%% file: main.tex
\documentclass{naturep}
\usepackage[utf8]{inputenc}
\usepackage{lineno}
\usepackage{setspace}
\usepackage{color}
\usepackage{csquotes}  
\usepackage{makecell}
\usepackage{fontawesome}
\usepackage[table]{xcolor}
\usepackage{colortbl}
\usepackage{graphicx}
\usepackage{booktabs}
\usepackage{amssymb}
\usepackage{amsmath}
\usepackage{hyperref}
\usepackage[T1]{fontenc}
\usepackage{tcolorbox}
\usepackage{mdframed}
\usepackage{adjustbox}
\usepackage{subcaption}
\usepackage{tablefootnote}
\usepackage[margin=0.6in]{geometry}
\usepackage{verbatim}
\usepackage{caption}
\usepackage{paralist}
\usepackage{cleveref}
\usepackage{multirow}
\usepackage{tabularx} 
\usepackage{array} 
\usepackage{float}
\usepackage{appendix}
\usepackage{hyperref} 
\usepackage[utf8]{inputenc}
\usepackage{geometry}
\usepackage{booktabs}
\usepackage{longtable} 
\usepackage{array}
\usepackage[table,xcdraw]{xcolor}
\usepackage{ragged2e}
\usepackage{setspace}
\usepackage{comment}

\definecolor{headerblue}{RGB}{230, 240, 255} 
\definecolor{rowgray}{RGB}{245, 245, 245}
\definecolor{darkblue}{RGB}{0, 50, 100}

\footnotesize
\definecolor{headerblue}{RGB}{220,230,242}
\definecolor{rowgray}{gray}{0.97}
\definecolor{highlightrow}{RGB}{255,245,210}

\newcommand\Heading[1]{
  \noindent\textbf{\Large{#1}}
}

\newcommand\heading[1]{
  \noindent\textbf{\large{#1}}
}

\newcommand\hheading[1]{
  \noindent\textbf{#1}
}

\makeatletter
\let\saved@includegraphics\includegraphics
\AtBeginDocument{\let\includegraphics\saved@includegraphics}

\makeatother

\title{Computational Pathology in the Era of Emerging Foundation and Agentic AI
\textit{--- International Expert Perspectives on Clinical Integration and Translational Readiness}}
\begin{document}
\maketitle
\begin{spacing}{1.2}

\author{
\noindent \noindent Qian Da$^{1,\textbf{*}}$, Yijiang Chen$^{2,\textbf{*}}$, Min Ju$^{3,\textbf{*}}$, Zheyi Ji$^{4,\textbf{*}}$, Albert Zhou$^{5,\textbf{*}}$,  Wenwen Wang$^{6,\textbf{*}}$, Matthew A Abikenari$^{7}$,  Philip Chikontwe$^{8}$, Guillaume Larghero$^{8}$, Bowen Chen$^{9}$, Peter Neidlinger$^{10}$,  Dingrong Zhong$^{11}$, Shuhao Wang$^{11}$, Wei Xu$^{12}$,  Drew Williamson$^{13}$, German Corredor$^{14}$, Sen Yang$^{15}$, Le Lu$^{15}$, Xiao Han$^{16}$, Kun-Hsing Yu$^{8,17, 18}$, Junzhou Huang$^{19}$, Laura Barisoni$^{20,21}$, Geert Litjens$^{22}$, Anant Madabhushi$^{15}$, Lifeng Zhu$^{23,\textbf{+}}$, Chaofu Wang$^{1,\textbf{+}}$, Junhan Zhao$^{8, 24, 25,\textbf{+}}$, Weiguo Hu$^{23,\textbf{+}}$
}
\end{spacing}

\vspace{-6mm}
\begin{spacing}{1.1}
\begin{affiliations}\small
\item{Department of Pathology, Ruijin Hospital, Shanghai Jiao Tong University School of Medicine, Shanghai, China}
\item{Department of Radiation Oncology, Stanford University School of Medicine, Stanford, CA, USA}
\item{Department of Basic Education, Qingdao City University, Qingdao, Shandong, China}
\item{Department of Human Genetics, the University of Chicago, Chicago, IL, USA}
\item{Department of Computer Science, University of Warwick, Coventry UK}
\item{Department of Electrical and Computer Engineering, Carnegie Mellon University, Pittsburgh, Pennsylvania, USA}
\item {Department of Neurosurgery, Stanford University School of Medicine, Stanford, CA, USA}
\item{Department of Biomedical Informatics, Harvard Medical School, Boston, MA, USA}
\item{Department of Biomedical Data Science, Stanford University, Stanford, CA, USA}
\item{Else Kroener Fresenius Center for Digital Health, Faculty of Medicine and University Hospital, TUD Dresden University of Technology, Dresden, Germany}
\item{Department of Pathology, China-Japan Friendship Hospital, Beijing, China}
\item{Institute for Interdisciplinary Information Sciences, Tsinghua University, Beijing, China}
\item{Department of Pathology and Laboratory Medicine, Emory University School of Medicine, Atlanta, GA, USA}
\item{Wallace H. Coulter Department of Biomedical Engineering, Georgia Institute of Technology and Emory University, Atlanta, GA, USA}
\item{Ant Group}
\item{College of Biomedical Engineering, Sichuan University, Sichuan, China.}
\item{Department of Pathology, Brigham and Women’s Hospital, Boston, MA, USA.}
\item{Harvard Data Science Initiative, Harvard University, Cambridge, MA, USA.}
\item{Department of Computer Science and Engineering, The University of Texas at Arlington, Arlington, TX, USA}
 \item{Department of Pathology, Division of AI and Computational Pathology, Duke University, Durham, NC, USA}
 \item{Department of Medicine, Division of Nephrology, Duke University, Durham, NC, USA}
\item{Computational Pathology Group, Oncode Institute, Department of Pathology, Radboudumc, Nijmegen 6525 GA, Netherlands}
\item{Shanghai Digital Medicine Innovation Center, Ruijin Hospital Affiliated to Shanghai Jiaotong University School of Medicine, Shanghai, China.}
\item{Department of Pediatrics, the University of Chicago, Chicago, IL, USA}
  \item{Comprehensive Cancer Center, The University of Chicago Medicine, Chicago, IL, USA}
\vspace{3mm}
 \\ \textbf{*} These authors contributed equally to this work.
 \\ \textbf[+]{Correspondences to: W.Hu, J.Zhao, C.Wang, and L.Zhu}
 \end{affiliations}
\end{spacing}

\newcommand{\PC}[1]{{\textcolor[rgb]{0,0,0}{#1}}}
\newcommand{\RE}[1]{{\textcolor[rgb]{1,0,0}{#1}}}
\newcommand{\JM}[1]{{\textcolor[RGB]{138,43,226}{#1}}}
\maketitle

\let\WriteBookmarks\relax
\def\floatpagepagefraction{1}
\def\textpagefraction{.001}

\clearpage
\Heading{Abstract}
\input{sections/0-abstract}

\clearpage
\begin{spacing}{1.35}
\Heading{1. Introduction}
\vspace{-1em}

\input{sections/1-introduction}

\Heading{2. Specialization and Generalization: Complementary Models in Computational Pathology}
\vspace{-1em}

\input{sections/2-specialization_generalization}



\Heading{3. Challenges to Translation: Economic and Technical Barriers}
\vspace{-1em}

\input{sections/3-translation_gap}


\Heading{4. Roadmap to Accelerated Clinical Adoption}

\input{sections/4-conclusion}
\vspace{1em}
\Heading{5. Conclusion}
\vspace{-1em}

\input{sections/5-perspective}

\end{spacing}

\newpage
\begin{tcolorbox}[colframe=blue!10!white, colback=blue!3!white, coltitle=black, title=Glossary, width=0.9\textwidth]
\setstretch{0.8}

\textbf{Clinical Concepts}
\begin{itemize}
    \item \textbf{Pathology Imaging and Diagnostic Techniques}
    \begin{itemize}
        \item \textbf{H\&E (Hematoxylin and Eosin)}: A widely used staining method in pathology to visualize tissue structures under a microscope.
        \item \textbf{IHC (Immunohistochemistry)}: A technique to identify specific antigens in tissues using antibodies, widely applied in cancer diagnosis.
        \item \textbf{WSI (Whole Slide Imaging)}: A digital scanning technique converting glass slides into high-resolution digital images.
        \item \textbf{ECT (Endoscopic Capsule Technology)}: A non-invasive diagnostic method using a capsule camera to capture images of the gastrointestinal tract.
    \end{itemize}
    
    \item \textbf{Clinical Applications and Support Systems}
    \begin{itemize}
        \item \textbf{Pathology Report Generation}: Automatically generating diagnostic reports from pathology images.
        \item \textbf{Cross-modal Analysis}: Analyzing multiple data types (e.g., combining text with images) for predictions.
        \item \textbf{Clinical Decision Support Systems (CDSS)}: AI-based systems assisting healthcare providers in clinical decisions.
        \item \textbf{Personalized Medicine}: Tailoring treatment based on individual genetic profiles, environment, and lifestyle.
    \end{itemize}
\end{itemize}

\vspace{0.5em}

\textbf{Technical Concepts}
\begin{itemize}
    \item \textbf{Model Architectures}
    \begin{itemize}
        \item \textbf{Transformers}: A deep learning architecture excelling in handling sequential data, particularly in natural language processing.
        \item \textbf{Foundation Models(FMs):} Large-scale deep learning models pretrained on vast datasets (often using self-supervision) that can be adapted to a wide range of downstream tasks.
        \item \textbf{Task-Specific Models (TSMs)}: Deep learning models designed and trained for a single, well-defined purpose, such as tumor detection or nuclei segmentation, lacking the broad generalization capabilities of foundation models.
        \item \textbf{Large Language Models (LLMs)}: Pretrained deep learning models designed for processing and generating human-like text.
    \end{itemize}
    
    \item \textbf{Learning Paradigms}
    \begin{itemize}
        \item \textbf{Multimodal Learning}: Integrating data from multiple modalities (e.g., text, image, sound) for robust analysis.
        \item \textbf{Pretrained Models}: Models trained on large datasets, fine-tuned for specific tasks.
        \item \textbf{Zero-shot Learning}: Enabling models to make predictions on tasks they have never seen during training.
        \item \textbf{Self-supervised Learning (SSL)}: A machine learning method where the model learns useful representations by predicting part of the data from other parts, without needing manually labeled data.
        \item \textbf{Few-shot Learning}: Generalizing from a small number of labeled examples.
        \item \textbf{Data Augmentation}: Increasing data diversity by modifying original data (e.g., rotating, flipping images).
    \end{itemize}
\end{itemize}

\end{tcolorbox}

\newpage
\Heading{References}
\bibliographystyle{nature}
\bibliography{used_ref}



\begin{appendices}

\input{sections/5-appendix}  
\end{appendices}

\end{document}

%% file: sections/0-abstract.tex
\begin{spacing}{1.2}
\noindent
\textbf{Recent breakthroughs in artificial intelligence through foundation models and agents have accelerated the evolution of computational pathology. Demonstrated performance gains reported across academia in benchmarking datasets in predictive tasks such as diagnosis, prognosis, and treatment response have ignited substantial enthusiasm for clinical application. Despite this development momentum, real world adoption has lagged, as implementation faces economic, technical, and administrative challenges. Beyond existing discussions of technical architectures and comparative performance, this review considers how these emerging AI systems can be responsibly integrated into medical practice by connecting deployable clinical relevance with downstream analytical capabilities and their technical maturity, operational readiness, and economic and regulatory context. Drawing on perspectives from an international group, we provide a practical assessment of current capabilities and barriers to adoption in patient care settings.}
\end{spacing}
\newpage

%% file: sections/1-introduction.tex

\noindent Histopathological evaluation of tissue remains the primary basis for cancer diagnosis and the starting point for nearly all downstream oncologic decision making, providing indispensable diagnostic, prognostic, and therapeutic guidance\cite{vanderlaak2021deep, madabhushi2019artificial}. As pathology increasingly informs prognostic stratification \cite{jiang2024end,wang2025foundation,armstrong2025development} and companion diagnostics \cite{Campanella2019_clinicalPathology, Virchow}, routine histopathological interpretation has become a central interface through which morphologic, molecular, and clinical information are combined to support therapeutic decisions \cite{PathChat, MUSK, valanarasu2026multimodal}. Computational pathology has emerged through the digitization of whole slide images (WSIs) \cite{2019Zarella_S_ArchivesofPathandLabMed_WSIPractice}, which enables machine learning systems to analyze histologic data acquired in routine clinical workflows \cite{vanderlaak2021deep, Campanella2019_clinicalPathology}. In step with advances in general artificial intelligence \cite{bommasani2021opportunities, moor2023foundation, kirillov2023segment}, computational pathology has progressed to foundation models (FMs) \cite{wang2022Ctranspath,UNI,CONCH,CHIEF,Virchow,GigaPath}, and is now beginning to move toward clinical AI agents \cite{ferber2025development, chen2025evidence, truhn2026artificial}.

\noindent Over the past decade, vision-only Task-Specific Models (TSMs) have enabled automated analysis, such as nuclei segmentation \cite{CoNSeP}, tumor detection \cite{Camelyon16}, and grade classification \cite{PANDA, wang2023generalizable}.These task specific models were typically optimized for individual applications, limiting their ability to generalize across tasks or institutions\cite{Campanella2019_clinicalPathology}. FMs overcome this bottleneck by leveraging large-scale representation learning on gigapixel WSIs through combinations of self-supervised learning, weak supervision, and knowledge distillation \cite{UNI, Ma2025GPFM}, enabling the extraction of robust features that generalize across diverse downstream applications without requiring extensive task-specific retraining \cite{Virchow, GigaPath}.These models have demonstrated performance in downstream analytical tasks, including diagnostic retrieval, molecular inference, risk modeling, and multimodal integration~\cite{MUSK,Virchow}.

\noindent Despite demonstrated performance gains on academic benchmarking datasets across predictive tasks, routine clinical adoption has lagged behind \cite{00aggarwal2025artificial, vanderlaak2021deep}.  Translating these emerging AI systems from \textit{in silico} validation to deployment in patient care settings faces a substantial ``reality gap,'' as performance metrics alone do not capture the economic, administrative, technical, and governance constraints of the clinical environment \cite{altucci2025artificial, xu2025discovering, vanderlaak2021deep}.Consequently, clinical deployment also requires governance frameworks to address safety, liability, diagnostic reliability, and economic sustainability\cite{truhn2026artificial,feng2025not,azad2026principles}. Recent reviews have detailed technical comparisons and perspectives on computational pathology models \cite{li2025multi, he2024foundation,truhn2026artificial}. This review extends the analysis beyond performance metrics to address the translational considerations necessary for clinical integration. Based on consensus among international experts in computational and clinical pathology, we establish a taxonomy to evaluate the utility of these models and examine the economic and regulatory conditions required to translate development momentum into deployable tools for patient care.

%% file: sections/2-specialization_generalization.tex
\noindent The integration of FMs into clinical pathology marks a transition from narrow, task-specific tools to generalizable diagnostic frameworks \cite{Virchow, PathChat}. Beyond mere automation, these architectures address fundamental challenges in manual assessment, such as high inter-observer variability and the cognitive load associated with complex case reviews \cite{bakas2024artificial, chew2024standardization}. By distilling high-dimensional morphological data into actionable insights, FMs serve as a scalable bridge between traditional histology and precision oncology, ensuring that routine workflows are both reproducible and adaptable to the increasing complexity of modern cancer diagnostics \cite{Campanella2025RealWorldDeployment} (\hyperlink{supp:A1}{Details in Extended Section A.1}).

\noindent While the emergence of Foundation Models represents a significant technological leap, it is critical to contextualize their utility against established  TSMs. TSMs continue to offer competitive performance with significantly lower computational overhead, particularly in scenarios where labeled data is abundant and clinical tasks are narrowly defined. We do not posit that FMs are universally superior; rather, we argue that their primary value lies in overcoming the data-efficiency bottlenecks that have historically constrained supervised learning. Therefore, the following sections provide our perspective on how FMs are re-shaping computational pathology not by rendering TSMs obsolete, but by offering a complementary paradigm essential for addressing the complexity and heterogeneity of modern oncology.

\vspace{-1em}
\section*{2.1 Augmenting Diagnostic Precision: From Routine Workflows to Rare Disease Retrieval}
\vspace{-1em}

FMs have fundamentally enhanced the precision of routine oncology workflows by capturing the hierarchical complexity of tissue architecture across pan-cancer cohorts \cite{Virchow,UNI}. Similar to traditional supervised methods, architectures such as UNI \cite{UNI}, TITAN \cite{TITAN}, and MUSK \cite{MUSK} robustly automate primary diagnostic tasks, including tumor detection and histological grading. Nonetheless, the foundation models maintained high fidelity and performance through self-supervised pre-training in data-constrained conditions \cite{kleppe2018chromatin, greten2023biomarkers}. This generalization is further exemplified by models like UNI \cite{UNI} and Virchow \cite{Virchow}, which effectively mitigate common performance drops caused by tissue artifacts or staining variability, ensuring reliable classification across both common and infrequent histological subtypes \cite{chai2025impact}.

\noindent The most transformative clinical utility of FMs, however, lies in addressing the "long-tail" challenges of pathology, where rare disease categories often lack sufficient labeled data for traditional AI training. By leveraging vision-language alignment, models such as CONCH \cite{CONCH}, PLIP \cite{PLIP}, and GigaPath \cite{GigaPath} enable zero-shot classification and content-based image retrieval (CBIR) from vast digital archives. This capability allows pathologists, particularly in non-specialist settings, to instantly retrieve and compare reference cases for rare tumors based on morphological similarity \cite{TITAN, PRISM}. By aligning visual features with semantic medical knowledge, these frameworks act as a scalable expert consultation system, reducing diagnostic uncertainty and facilitating the rapid integration of emerging diagnostic categories into clinical practice \cite{TITAN, CONCH}(\hyperlink{supp:A2}{ Details in Extended Section A.2-A.4}).

\vspace{-1em}
\section*{2.2 The "Virtual Assay": Molecular Profiling and Spatial Integration}
\vspace{-1em}

\subsection{2.2.1 Molecular Biomarker Prediction}
FMs have demonstrated potential utility as "virtual assays" by directly inferring molecular signatures from H\&E-stained WSIs\cite{wang2024screen}. Models such as GigaPath \cite{GigaPath} and TITAN \cite{TITAN} robustly predict oncogenic alterations, including TP53, KRAS, and EGFR mutations, which critically influence treatment selection across multiple cancer types. Beyond single-gene mutations, these architectures effectively identify complex biomarkers such as tumor mutational burden (TMB) \cite{ GigaPath} and microsatellite instability (MSI) \cite{wagner2023transformer, PRISM}. By capturing sub-visual morphological correlates, such as tumor-infiltrating lymphocytes and specific glandular patterns, FMs provide a rapid, cost-effective alternative to traditional sequencing for identifying immunotherapy candidates \cite{wang2024screen}.

\noindent However, clinical translation requires rigorous control of confounding effects; for example, high predictive accuracy for \textit{BRAF} mutations in colorectal cancer may partly reflect their strong statistical association with MSI status rather than direct morphological manifestations \cite{wagner2023transformer}. Ensuring that these models reflect genuine biological relationships is essential for their reliability as diagnostic adjuncts. Beyond genomics, the scope of these image-based assays is expanding to include protein expression typically assessed by immunohistochemistry (IHC) \cite{hua2024pathoduet,MUSK}, as well as DNA methylation status and transcriptomic subtypes \cite{kather2024biomarker}. This integration of FMs into the molecular diagnostic pipeline offers a scalable pathway for real-time patient stratification in precision oncology.

\subsection{2.2.2 Virtual Spatial Biology}

FMs extend the "virtual assay" paradigm by aligning histological images with diagnostic language and spatially resolved molecular data. Through multimodal pretraining on large-scale datasets \cite{PathAlign, PathGen}, architectures such as MUSK\cite{MUSK}, PRISM \cite{PRISM} and TITAN \cite{TITAN} process gigapixel-scale images to generate coherent diagnostic narratives and clinical reports without manual annotations. To ensure granular accuracy for staging and treatment planning, models like MI-Gen \cite{MI_Gen} utilize multi-instance learning to capture localized micro-features (\hyperlink{supp:A4}{Extended Section A.4}). This vision-language alignment further enables bidirectional retrieval and interactive reasoning; for example, CONCH \cite{CONCH} and MUSK \cite{MUSK} facilitate natural language-based case searches, while PathChat \cite{PathChat} integrates patient history into a conversational interface for dynamic diagnostic support.

\noindent Beyond text, FMs increasingly integrate histology with spatial transcriptomics (ST) and spatial proteomics (SP) \cite{chen2025OmiCLIP,li2026ai} to learn the "morphological language" of molecular expression \cite{hao2024scFoundation, gulati2025spatialtrans}. Utilizing curated image-transcriptome datasets such as HEST-1k \cite{jaume2024hest}, frameworks like scGPT \cite{wang2025scgpt}, scFoundation \cite{hao2024scFoundation}, and OmiCLIP \cite{chen2025OmiCLIP} link tissue architecture to spatially resolved gene profiles. Similarly, in the protein domain, HEX \cite{li2026ai}(based on MUSK architecture), have demonstrated high-resolution patch level prediction of high-plex spatial protein expression from H\&E morphology, and KRONOS \cite{KRONOS} captures high-dimensional expression patterns at single-cell resolution. However, recent benchmarking indicates that predicting spatial gene expression from morphology alone remains translationally challenging due to inherent tissue heterogeneity \cite{wang2025scgpt}. Despite these limitations, these spatially grounded models represent a necessary evolution to disentangle complex biological signals that unimodal systems cannot resolve, providing a higher-resolution understanding of the tumor microenvironment.
\vspace{-1em}
\section*{2.3 Predictive and Prognostic Oncology: Risk Stratification and Companion Diagnostics}
FMs refine clinical decision-making by distilling high-dimensional morphological signals into actionable prognostic insights \cite{jiang2024end,MUSK, yuan2025pancancer}. Recent pan-cancer architectures, including TITAN \cite{TITAN}, BEPH \cite{BEPH}, and Prov-GigaPath \cite{ wang2025foundation}, identify complex survival determinants such as cellular atypia, architectural fragmentation, and stromal remodeling that often escape manual assessment. By integrating these histological features with clinical metadata, multimodal FMs quantify risks of tumor recurrence and long-term therapeutic trajectories with higher granularity than conventional TNM staging \cite{MUSK,yuan2025pancancer}.

\noindent In the domain of immunotherapy, FMs function as essential companion diagnostics by uncovering sub-visual indicators of immune evasion. Where traditional biomarkers like PD-L1 often fail, particularly in \textit{EGFR}-mutant or cold tumors, models such as MUSK \cite{MUSK} reliably identify responders by characterizing the spatial distribution of tumor-infiltrating lymphocytes (TILs) and tertiary lymphoid structures. Furthermore, FMs are increasingly utilized to model neoadjuvant therapy responses, quantifying longitudinal changes in the tumor microenvironment (TME) such as vascular regression and therapy-induced fibrosis \cite{vaidya2025molecular}. This predictive precision is transforming clinical trial design, enabling pharmaceutical collaborations to optimize enrollment through AI-driven patient selection, thereby reducing trial attrition and accelerating the delivery of targeted and neoadjuvant regimens \cite{arango2025ai_clinical_enrollment, harrer2024artificial, mandl2025ai}.

\vspace{-1em}

\section*{2.4 The AI Assistant: Multimodal Synthesis and Agentic Reasoning}
\vspace{-1em}

\subsection{2.4.1 The Digital Multidisciplinary Tumor Board}
Diagnostic reasoning in oncology requires the synthesis of histology, clinical history, and molecular profiles to construct a comprehensive patient profile. FMs are evolving to automate this integration, effectively functioning as "Digital Multidisciplinary Tumor Boards" (MTBs). By utilizing multimodal pretraining on diverse datasets, such as the image–text pairs in PathAlign \cite{PathAlign} and PathGen \cite{PathGen}, these models learn a shared semantic space that aligns histological visual features with diagnostic language and clinical metadata.

\noindent To translate this into clinical utility, slide-level architectures like PRISM \cite{PRISM}, TITAN \cite{TITAN}, and Prov-GigaPath \cite{GigaPath} aggregate gigapixel-scale information to cross-reference morphological patterns with genomic signatures or survival outcomes. This global perspective allows the AI to ground abstract visual features in molecular truth, uncovering correlations invisible to manual assessment. Furthermore, models like MI-Gen \cite{MI_Gen} and CONCH \cite{CONCH} employ multi-instance learning and contrastive frameworks to integrate localized micro-features—such as focal invasion—with broader clinical context. This multimodal synthesis is increasingly interactive; tools like PathChat \cite{PathChat} and MUSK \cite{MUSK} enable pathologists to query complex cases via natural language, incorporating patient history and domain-specific terminology to navigate difficult diagnoses dynamically. Collectively, these innovations streamline established workflows by reducing administrative burden while ensuring that diagnostic narratives are both granular and consistent across specialized and non-specialized centers \cite{MI_Gen, PathChat}.

\subsection{2.4.2 Generative Reporting}
The integration of textual modalities shifts AI from a "black-box" classifier toward an interpretable assistant capable of automated report generation. To bridge the gap between gigapixel-scale WSIs, architectures like GIGAPATH\cite{GigaPath}, WSI-LLaVA\cite{liang2025wsi}, PRISM \cite{PRISM} and TITAN \cite{TITAN} leverage self-supervised encoders and transformer-style decoders to compress thousands of patches into compact slide-level representations. On top of that, PRISM\cite{PRISM} and TITAN\cite{TITAN} also integrated such representations with clinical language. This enables the autoregressive generation of coherent diagnostic narratives without manual region-of-interest annotations, while maintaining a global understanding of tissue context.

\noindent To ensure clinical relevance, these frameworks must capture both holistic patterns and rare, localized findings. Models such as MI-Gen \cite{MI_Gen} adopt a multi-instance learning paradigm that avoids simplistic slide-level embeddings. Instead, an encoder-decoder architecture allows the language-model decoder to dynamically query specific patch-level features through an attention mechanism (\hyperlink{supp:A4}{Details in Extended Section A.4}). This granular approach ensures that critical micro-features, such as a single focus of lymphovascular invasion, are integrated into the generated text, supporting accurate staging and treatment planning. While achieving expert-level precision in quantitative reporting remains an open challenge \cite{Tran2025generatingDermatopathology}, applying scaling laws to diverse pathological corpora \cite{kaplan2020scaling} provides a pathway to streamline workflows and improve inter-observer consistency in routine documentation.

\subsection{2.4.3 From Chatbots to Agentic Reasoning}
While conversational assistants like PathChat\cite{PathChat} empower pathologists to interrogate slides via passive question and answer, the field is advancing toward Agentic AI\cite{sun2025cpathagent, chen2025pathagent}, typified by systems like SlideSeek \cite{chen2025evidence}. Unlike static classifiers, agentic systems utilize a hierarchical "Supervisor-Explorer" architecture that mimics the human cognitive workflow\cite{sun2025cpathagent, chen2025pathagent}. A "Supervisor" agent formulates a diagnostic hypothesis and autonomously directs "Explorer" agents to scan the slide, zooming in on regions of interest (ROI) to gather confirmatory evidence. This iterative process generates an interpretable reasoning chain, explicitly linking diagnostic conclusions to the specific ROI coordinates visited.  Such implementations will translate foundation models from passive tools into transparent partners that demonstrate their logic\cite{truhn2026artificial}.

\begin{figure*}[!ht]
\setstretch{0.9}
  \centering
  \includegraphics[width=\linewidth]{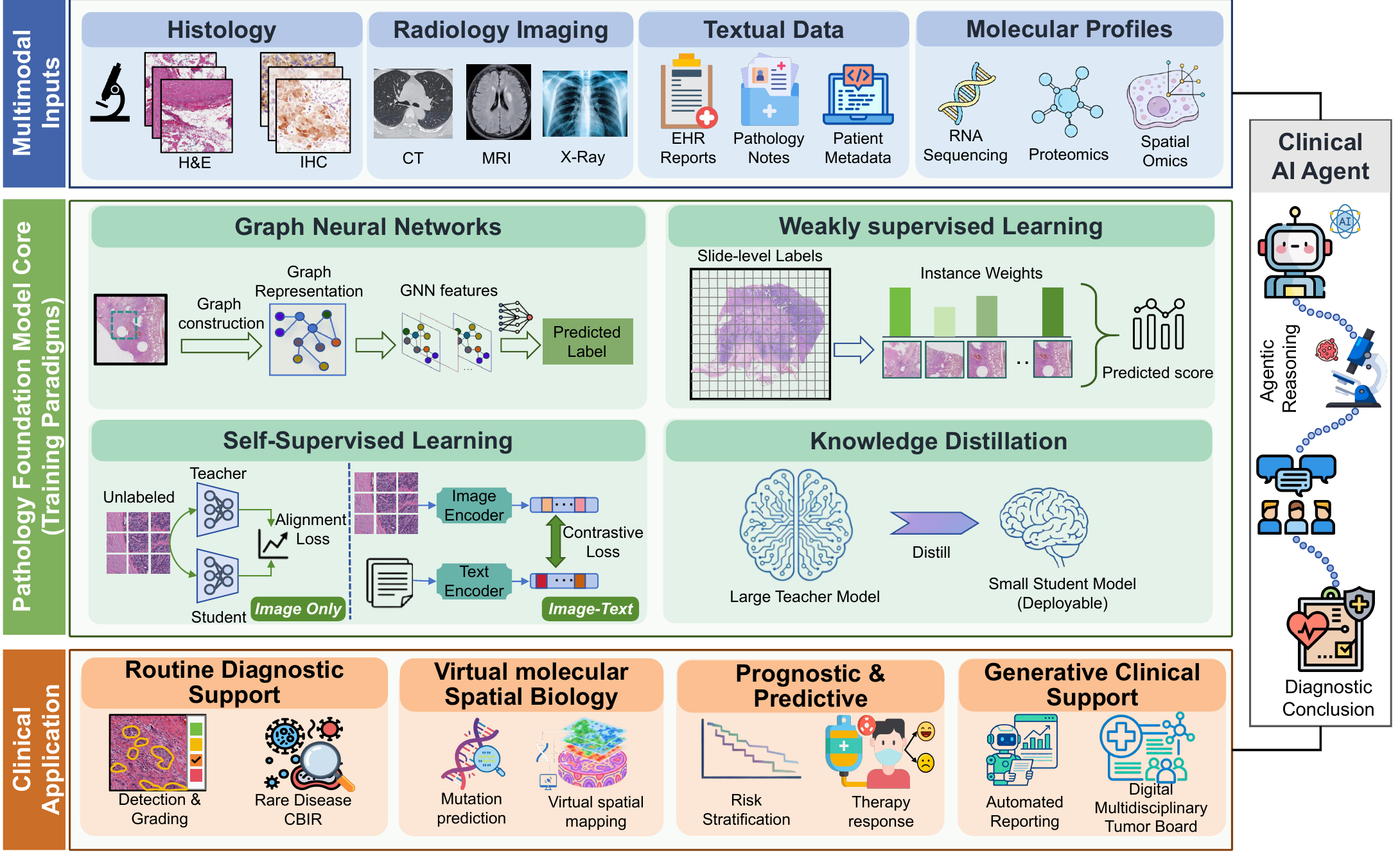}
  \caption{\textbf{The evolution of multimodal pathology foundation models from passive diagnostic aids to autonomous clinical orchestrators.} This framework represents a paradigm shift in computational pathology\cite{campanella2024computational}, where the integration of disparate data streams\cite{seoni2024all, Trident} enables models to transcend traditional morphological assessment and ground abstract visual features in molecular truth\cite{kather2024biomarker, gulati2025spatialtrans}. By leveraging self-supervised alignment and structural modeling\cite{chen2022fast, UNI}, these systems bridge the gap between gigapixel-scale imaging and actionable prognostic insights, uncovering sub-visual determinants of survival and immune evasion that often escape manual evaluation\cite{zimmermann2024virchow2}. The hierarchical application tiers reflect an increasing complexity of clinical utility, moving from routine phenotypic characterization toward "virtual assays"\cite{AIVC, VirTues} that map the spatial and molecular landscape of the tumor microenvironment without additional tissue consumption. Ultimately, the transition toward agentic reasoning, facilitated by a "Supervisor-Explorer" architecture, transforms the AI from a black-box classifier into a transparent diagnostic partner that mimics expert cognitive workflows, providing an interpretable reasoning chain for complex clinical decision-making and personalized therapeutic planning\cite{ferber2025development, ghezloo2025pathfinder}.} 
\label{fig:workflow}
\end{figure*}
\vspace{-1em}

\begin{table*}[t]
\footnotesize
\setstretch{0.95}
\caption{\setstretch{0.95} Commonly used datasets in computational pathology, grouped by primary organ/scope.
Modality is specified as Whole Slide Images (WSI), microscopy images (field-of-view), or patches/ROIs to distinguish between acquisition scales.
Multi-modal datasets include additional data types such as radiology, genomics, or clinical records.}
\label{table:list_of_dataset_final}

\resizebox{\textwidth}{!}{%
\begin{tabular}{
  >{\columncolor{headerblue}\centering\arraybackslash}p{2.7cm}
  >{\columncolor{headerblue}\centering\arraybackslash}p{4.2cm}
  >{\columncolor{headerblue}\centering\arraybackslash}p{6.4cm}
  >{\columncolor{headerblue}\centering\arraybackslash}p{5.6cm}
  >{\columncolor{headerblue}\centering\arraybackslash}p{4.1cm}
}
\toprule
\textbf{Dataset} & \textbf{Organs / Cancer types} & \textbf{Modality} & \textbf{Annotation type} & \textbf{Dataset size} \\
\midrule

\rowcolor{headerblue!18}
\multicolumn{5}{l}{\textbf{Multi-organ consortia / multi-modal repositories}}\\

\rowcolor{rowgray}
HTAN &
Breast, Lung, Colorectal, and more (21 total) &
\makecell[l]{Multiplex tissue images,\\Clinical data, Genomics} &
Various (subtype, diagnosis) &
\makecell[l]{2,188 cases;\\9,143 biospecimens} \\

PLCO &
Prostate, Lung, Colorectal, Ovarian &
\makecell[l]{Histopathology slides,\\Chest X-ray, Biospecimens} &
Subtype, clinical data &
\makecell[l]{13,165 pathology images;\\X-rays available} \\

\rowcolor{rowgray}
CPTAC-TCIA &
Breast, Lung, Colon, and more &
\makecell[l]{Histopathology slides,\\Radiology (CT/MR/PET)} &
Various (subtype, stage) &
\makecell[l]{305,898 samples} \\

TCGA &
33 cancer types &
\makecell[l]{Histopathology WSI,\\Genomics, Clinical} &
Various (subtype, stage) &
\makecell[l]{33,179 samples} \\

\rowcolor{rowgray}
CPIA &
48 organs, 100+ diseases &
\makecell[l]{Histopathology images\\(WSI \& ROI)} &
Subtype; multi-scale labels &
\makecell[l]{21,427,877 images} \\

PathMMU &
Breast, Lung, Liver, Colon, and more &
\makecell[l]{Histopathology patches,\\QA pairs} &
Cancer type; diagnostic description &
\makecell[l]{24,067 patches} \\

\rowcolor{rowgray}
MIDOG++ &
Multi-organ (Breast, Lung, Lymphoma, etc.) &
\makecell[l]{Histopathology ROIs} &
Mitotic figure detection &
\makecell[l]{503 tumor cases} \\

PanNuke &
19 cancer types &
\makecell[l]{Histopathology nuclei tiles} &
Nuclei instance masks; class (5) &
\makecell[l]{189,744 nuclei instances} \\

\rowcolor{rowgray}
MoNuSAC2020 &
Lung, Prostate, Kidney, Breast &
\makecell[l]{Histopathology nuclei tiles} &
Instance segmentation; nuclei class &
\makecell[l]{31,411 nuclei instances} \\

\rowcolor{rowgray}
ARCH &
Various cancer types &
\makecell[l]{Histopathology patches,\\Text captions} &
Dense captions &
\makecell[l]{4,270 patches} \\

SegPath &
Various cancer types &
\makecell[l]{Histopathology patches} &
Semantic segmentation masks &
\makecell[l]{158,687 patches} \\

\midrule

\rowcolor{headerblue!18}
\multicolumn{5}{l}{\textbf{Breast (incl. lymph node metastasis)}}\\

\rowcolor{rowgray}
Camelyon16 &
Breast (lymph node) &
\makecell[l]{Histopathology WSI} &
Metastasis / no metastasis &
\makecell[l]{400 WSIs} \\

Camelyon17 &
Breast (lymph node) &
\makecell[l]{Histopathology WSI (H\&E)} &
Metastasis / no metastasis &
\makecell[l]{1,000 WSIs} \\

\rowcolor{rowgray}
PCam &
Lymph nodes &
\makecell[l]{Histopathology patches} &
Metastasis / no metastasis &
\makecell[l]{327,680 patches} \\

BreaKHis &
Breast &
\makecell[l]{Microscopy images} &
Benign/malignant; tumor type (8) &
\makecell[l]{7,909 images} \\

\rowcolor{rowgray}
BRACS &
Breast lesions &
\makecell[l]{Histopathology WSI + ROIs} &
Benign/malignant; 6 subtypes &
\makecell[l]{547 WSIs; 4,539 ROIs} \\

BACH &
Breast &
\makecell[l]{Microscopy images} &
4-class (normal/benign/in situ/invasive) &
\makecell[l]{400 images} \\

\rowcolor{rowgray}
BCSS &
Breast &
\makecell[l]{Histopathology WSI + patches} &
Segmentation mask &
\makecell[l]{151 WSIs; 20,000 patches} \\

BreCaHAD &
Breast &
\makecell[l]{Histopathology patches} &
Centroids + 6 labels &
\makecell[l]{162 patches} \\

\rowcolor{rowgray}
TUPAC16 &
Breast &
\makecell[l]{Histopathology WSI} &
Class (WSI-level) &
\makecell[l]{500 WSIs} \\

HEROHE-ECDP2020 &
Breast &
\makecell[l]{Histopathology WSI} &
Positive/negative &
\makecell[l]{509 WSIs} \\

\rowcolor{rowgray}
NuCLS &
Breast &
\makecell[l]{Histopathology ROIs} &
Nuclear detection; nuclei class &
\makecell[l]{3,944 ROIs; 220,000 nuclei} \\

SLN-Breast &
Breast &
\makecell[l]{Histopathology WSI} &
Cancer / no cancer &
\makecell[l]{130 WSIs} \\

\rowcolor{rowgray}
TIGER &
Breast &
\makecell[l]{Histopathology WSI + ROIs} &
Segmentation; TILs score &
\makecell[l]{390 WSIs (approx);\\+ ROI subsets} \\

\midrule

\rowcolor{headerblue!18}
\multicolumn{5}{l}{\textbf{Colorectal / GI}}\\

\rowcolor{rowgray}
NCT-CRC-HE-100K &
Colorectal &
\makecell[l]{Histopathology patches (H\&E)} &
Cancer type &
\makecell[l]{100,000 patches} \\

EBHI-Seg &
Colorectal &
\makecell[l]{Histopathology patches} &
Segmentation masks &
\makecell[l]{5,710 patches} \\

\rowcolor{rowgray}
MHIST &
Colorectal polyps &
\makecell[l]{Histopathology patches} &
Benign/malignant &
\makecell[l]{3,152 patches} \\

CoNSeP &
Colorectal cancer &
\makecell[l]{Histopathology tiles} &
Nuclei segmentation &
\makecell[l]{41 tiles} \\

\rowcolor{rowgray}
GlaS &
Colorectal (gland) &
\makecell[l]{Histopathology patches} &
Benign/malignant; gland labels &
\makecell[l]{165 patches} \\

UniToPatho &
Colon &
\makecell[l]{Histopathology patches} &
Class (6) &
\makecell[l]{9,536 patches} \\

\rowcolor{rowgray}
HunCRC &
Colon &
\makecell[l]{Histopathology patches + WSI} &
Class (10) &
\makecell[l]{101,389 patches; 200 WSIs} \\

\midrule

\rowcolor{headerblue!18}
\multicolumn{5}{l}{\textbf{Prostate}}\\

\rowcolor{rowgray}
SegPANDA200 &
Prostate &
\makecell[l]{Histopathology patches} &
Gleason grade &
\makecell[l]{100,960 patches} \\

SICAPv2 &
Prostate cancer &
\makecell[l]{Histopathology patches} &
Gleason scores/grades &
\makecell[l]{18,783 patches} \\

\rowcolor{rowgray}
RINGS &
Prostate &
\makecell[l]{Histopathology WSI} &
Gland/tumor segmentation &
\makecell[l]{1,500 WSIs} \\

\midrule

\rowcolor{headerblue!18}
\multicolumn{5}{l}{\textbf{Lung}}\\

\rowcolor{rowgray}
WSSS4LUAD &
Lung (LUAD) &
\makecell[l]{Histopathology patches} &
Tumor / stroma / normal &
\makecell[l]{10,211 patches} \\

LubLung &
Lung &
\makecell[l]{Histopathology patches} &
Class (9) &
\makecell[l]{23,199 patches} \\

\midrule

\rowcolor{headerblue!18}
\multicolumn{5}{l}{\textbf{Kidney / Skin / Ovary}}\\

\rowcolor{rowgray}
RenalCell &
Kidney &
\makecell[l]{Histopathology patches} &
Subtypes / benign vs malignant &
\makecell[l]{64,553 patches} \\

SkinCancer &
Skin tumors &
\makecell[l]{Histopathology patches} &
Benign/malignant/other &
\makecell[l]{129,364 patches} \\

\rowcolor{rowgray}
Ovarian Bevacizumab Response &
Ovary &
\makecell[l]{Histopathology WSI,\\Clinical data} &
Treatment effectiveness &
\makecell[l]{288 WSIs} \\

\midrule

\rowcolor{headerblue!18}
\multicolumn{5}{l}{\textbf{Mixed organ}}\\

\rowcolor{rowgray}
LC25000 &
Lung / Colon cancer &
\makecell[l]{Histopathology patches} &
Cancer type (5) &
\makecell[l]{25,000 patches} \\

\bottomrule
\end{tabular}}
\end{table*}

%% file: sections/3-translation_gap.tex
{\color{blue}}


\noindent Despite the celebration of academic success in developing and validating pathology FMs\cite{he2024foundation, GigaPath, CHIEF, UNI}, reporting expert-level performance across diverse diagnostic tasks, only a small fraction of these systems have reached sustained deployment in hospital settings\cite{Campanella2025RealWorldDeployment, vanderlaak2021deep}. This disconnect reflects not a transient implementation delay, but a deeper structural misalignment between the incentives that govern academic innovation and the practical requirements of clinical adoption. Dominant research paradigms prioritize benchmark performance on curated datasets, architectural novelty, and short-term retrospective validation, yet only weakly capture the longitudinal stability, workflow compatibility, and risk management demanded by clinical environments\cite{El_Arab2025, vasey2022reporting, reinke2024understanding}.

\noindent Unlike pharmaceutical agents or medical devices that enter well-established regulatory, reimbursement, and liability frameworks, AI-based pathology systems occupy a fragmented translational landscape. Economic viability, technical robustness, institutional integration, and legal accountability remain only partially resolved\cite{stern2023economics, price2024liability}, rendering many scientifically sophisticated models operationally fragile\cite{de2025current, Howard2021SiteSpecificSignatures}. Although pharmaceutical and biotechnology industries have increasingly adopted pathology AI for biomarker discovery and trial optimization\cite{harrer2024artificial, arango2025ai_clinical_enrollment}, these vertically integrated, sponsor-funded applications operate in controlled environments and do not address the systemic barriers faced by hospital laboratories. Here, we rethink the pathology AI translation gap across economic, technical, and sociotechnical dimensions, and argue that durable clinical impact will require a shift from dataset-centric optimization toward system-level design principles that prioritize robustness, accountability, economic sustainability, and institutional alignment.

\vspace{-1em}

\section*{3.1 Economic Reality: Development  Deployment, and Reimbursement}
\vspace{-1em}

The economic barriers to clinical translation of pathology FMs extend beyond high absolute costs to a fundamental asymmetry in how financial risk is distributed across development, deployment, and long-term operation. Model development requires sustained investment in specialized AI expertise, large-scale computing infrastructure, curated training datasets, and extensive clinical annotation. These early-stage costs are typically supported through public research funding, venture capital, and pharmaceutical partnerships, enabling rapid technical progress while externalizing much of the initial financial risk from healthcare institutions\cite{stern2023economics, Baxi2022}.

\noindent In contrast, responsibility for data generation and digital infrastructure falls largely on hospitals and diagnostic laboratories. Unlike radiology, which standardized digital workflows decades ago, many pathology departments remain partially or fully analog\cite{lujan2021dissecting, evans2022regulatory}. Implementing AI therefore necessitates a comprehensive transition to digital pathology, requiring substantial capital investment in high-throughput whole-slide scanners and associated information systems (\hyperlink{supp:B1}{Extended Section B.1 and B.3}). Regulatory requirements further shape this cost structure, as FDA authorizations often mandate specific acquisition and processing pipelines, effectively constraining institutions to validated platforms from major vendors such as Philips, Leica, or Roche\cite{schumacher2021application, muralidharan2024scoping}. These constraints limit vendor competition and increase long-term procurement costs. Beyond hardware, large-scale digitization generates cascading computational and data management expenses\cite{verghese2023computational}. A medium-sized hospital in populated countries like China or India producing approximately 3,000 slides per day may accumulate petabyte-scale data annually, transforming digital pathology into a persistent operational burden\cite{mulliqi2025foundation, holub2023privacy}. This challenge is amplified by regulatory asymmetries in the United States. Under CLIA regulations, physical glass slides remain the legal primary record with mandatory long-term retention\cite{clia_retention}, while digital files lack unified federal preservation mandates. Institutions are therefore compelled to maintain parallel physical archives and cloud-based repositories, effectively doubling storage and compliance costs in the absence of corresponding reimbursement\cite{homeyer2022artificial}.

\noindent The financial burden intensifies during long-term deployment and maintenance. Sustained clinical use requires continuous model monitoring, periodic performance audits, cybersecurity protection, integration with evolving laboratory information systems, and recurrent regulatory revalidation\cite{El_Arab2025, gilbert2023algorithm}(\hyperlink{supp:B2}{Details in Extended Section B.2}). Unlike conventional laboratory instruments with predictable depreciation schedules, AI systems demand ongoing reinvestment to manage data drift, software updates, and changing clinical standards. These recurrent expenditures convert AI infrastructure from a one-time capital purchase into a persistent operational liability\cite{feng2025not}. Despite assuming the majority of downstream financial risk, healthcare institutions capture only limited direct economic return from these investments. Many potential benefits, such as improved workflow efficiency, reduced diagnostic delays, enhanced trial enrollment, and payer-side cost savings, are sometimes hard to measure and distributed diffusely across insurers, industry sponsors, and technology vendors\cite{gondi2023paying}. In the absence of mechanisms to internalize these external benefits, hospitals are left to absorb substantial costs while realizing uncertain financial gains. As a result, the economic calculus for large-scale adoption often disfavors sustained institutional investment, even when technical performance is well established\cite{Hassan2024}.

\noindent A central consequence of this cost–benefit asymmetry is the lack of sustainable reimbursement pathways. In the United States, the absence of permanent CPT codes for AI-assisted interpretation compels pathology departments to absorb implementation and maintenance expenses as unreimbursed operational overhead\cite{gondi2023paying, evans2022regulatory}. Expenditures are frequently justified through speculative gains in efficiency rather than guaranteed revenue streams, reinforcing conservative adoption strategies and limiting long-term planning. Global strategies illustrate different policy and pricing approaches. China has implemented a centralized pricing framework that integrates AI-assisted diagnostics into reimbursable medical services\cite{nhsa2025guide}. By explicitly linking digital infrastructure and algorithmic interpretation to billable activities, this model incentivizes hospital–industry collaboration and supports scalable deployment. The U.S. market has multiple commercial pathways, such as Laboratory Developed Tests (LDTs) and pharmaceutical-sponsored deployments\cite{spratt2023artificial, Kurnat_Thoma2024}. Although these models have shown early adoption by supporting product maturation, revenue generation and locally clinical deployment, they have often remained confined to pilot settings and have not fully generated system-wide impact or addressed the broader challenges of sustainability\cite{homeyer2022artificial}.

\vspace{-1em}

\begin{figure*}[!ht]
\setstretch{0.9}
  \centering
\includegraphics[width=\columnwidth]{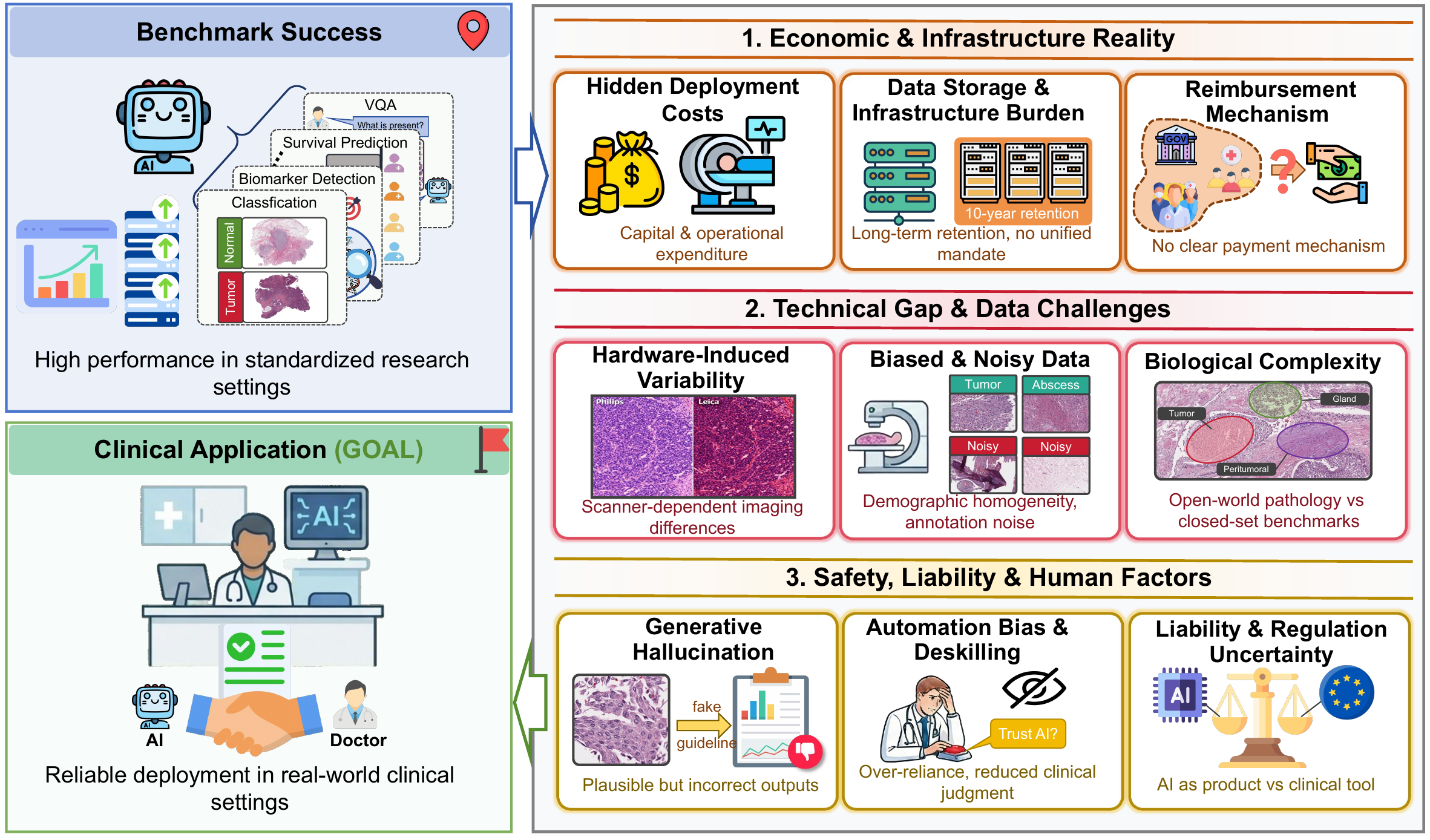}
\caption{\textbf{Translational Opportunities and Barriers for Clinical Implementation of Pathology FMs .}
We summarize three fundamental gaps that must be bridged to translate basic pathology AI models into real-world clinical application.
(1) \emph{Economic and infrastructure reality}, including indirect deployment costs, large-scale data storage, long-term maintenance of imaging infrastructure, and unresolved questions regarding reimbursement and sustainability.
(2) \emph{Technical gap and data challenges}, arising from hardware heterogeneity, scanner- and site-specific variability, data quality and bias risks, and the intrinsic biological complexity that limits robustness and generalizability.
(3) \emph{Safety, liability, and human factors}, encompassing risks of generative hallucinations, cognitive automation bias, clinician deskilling, and the need for evolving accountability, regulatory, and legal frameworks for clinical adoption.}

  \label{fig:gap}
\end{figure*}
\vspace{-1em}

\section*{3.2 The Reality Gap: Technical Barriers to Reliable Clinical Deployment}
\vspace{-1em}
Beyond economic constraints, a fundamental barrier to clinical translation lies in the persistent gap between the technical assumptions underlying pathology FMs  and the realities of routine diagnostic practice. Most contemporary systems are developed and validated under controlled conditions using curated datasets, standardized workflows, and narrowly defined tasks. In contrast, clinical pathology operates in heterogeneous, noisy, and evolving environments, where diagnostic reasoning integrates diverse sources of uncertainty. As a result, models that perform well on retrospective benchmarks may fail to meet the reliability and interpretability standards required for routine clinical use (\hyperlink{supp:B4}{Details in Extended Section B.4}). Another bottleneck is interoperability in deployment\cite{homeyer2022artificial, clunie2021dicom}; unlike radiology, which standardized workflows around DICOM decades ago, digital pathology remains fragmented by proprietary scanner formats that often lack native compatibility with Laboratory Information Systems\cite{Herrmann2018DICOM}. Consequently, hospitals cannot easily integrate model outputs into the pathologist's routine viewer without complex, custom middleware.

\noindent However, the most critical aspect of the technical and clinical reality gap is the pervasive domain shift driven by pre-analytic variabilities introduced throughout digital pathology data generation. These variations manifest across multiple parallel stages: First, hardware-induced shifts remain a primary factor; although major whole-slide scanners are FDA-cleared, they employ proprietary optical sensors, color calibration pipelines, and reconstruction algorithms. Consequently, the same glass slide digitized on different platforms can yield images with substantially different color spaces and contrast profiles\cite{tellez2019quantifying, Howard2021SiteSpecificSignatures}. Second, and equally impactful, is the lack of standardization in wet-lab protocols across pathology laboratories globally. Inconsistencies arise in every step of physical slide generation, including variations in tissue slide preparation, the thickness of the cut, stain intensity, and the presence of extraneous artifacts such as pen markings\cite{janowczyk2019histoqc, shah2025impact}. Third, the structural distinction between WSIs and Tissue Microarrays introduces fundamental differences in tissue representation and context, further complicating data homogeneity. Collectively, these factors create significant challenges for the calibration of FMs  in clinical deployment. When pretraining datasets lack diversity across these axes, models risk conflating technical batch effects with actual biological signals\cite{schmitt2021hidden}. This creates a dependency where robust deployment requires representations that decouple tissue morphology from this wide spectrum of generation-side variables, rather than just hardware characteristics alone\cite{UNI, li2025open}(\hyperlink{supp:B5}{Details in Extended Section B.5}).

\noindent The reliability of learned representations is further constrained by intrinsic limitations in pretraining data quality and representational bias. Large-scale datasets often contain heterogeneous annotation standards, ambiguous labels, and systematic noise. Recent studies have described the phenomenon of catastrophic inheritance, in which FMs  trained on noisily labeled corpora encode latent structural defects that persist through downstream fine-tuning\cite{sun2025from}. Even modest levels of upstream label corruption can distort feature space geometry in ways that are difficult to correct post hoc\cite{sun2025from}. These findings underscore that scaling data volume without rigorous curation may amplify, rather than mitigate, downstream clinical risk. Population and institutional biases further restrict generalizability. Many widely used training resources, including TCGA, are derived predominantly from populations of European ancestry and a limited set of academic medical centers\cite{spratt2016racial, vaidya2024demographic, fischer2024fairness}. Models trained on such cohorts may establish biased reference baselines, exhibiting degraded sensitivity in underrepresented populations or region-specific disease subtypes\cite{fischer2024fairness, vaidya2024demographic}. Moreover, site-specific artifacts can induce clustering by institution rather than biological class, reinforcing spurious correlations that undermine external validity\cite{de2025current}. Without large-scale, geographically and demographically diverse validation, such biases risk being propagated into clinical practice.

\noindent Broadly, similar to other AI-assisted vision tools, a critical gap still persists between benchmark-oriented evaluation paradigms and the cognitive demands of real-world diagnosis\cite{da2025survey, El_Arab2025}. Clinical pathology rarely conforms to closed-set classification tasks. Individual slides frequently contain complex mixtures of malignant, benign, inflammatory, and reactive processes. For example, identifying papillary thyroid carcinoma within nodular hyperplasia requires disentangling subtle neoplastic features from extensive benign background\cite{hong2022distinct}, while benign mimics such as microglandular adenosis may coexist with invasive ductal carcinoma in breast specimens\cite{lester2025diagnostic}. Human experts resolve such ambiguity through hierarchical reasoning and contextual integration. In contrast, current FMs, which largely lack explicit mechanisms for multiscale abstraction and causal interpretation, may conflate mimics with malignancy or fail to localize clinically decisive components\cite{sun2025cpathagent, wang2025pathology}(\hyperlink{supp:B6}{Details in Extended Section B.6}).To guide this translation, Table~\ref{tab:deployment_risk} presents a structured framework categorizing these deployment dimensions along with specific detection and mitigation strategies.

\begin{table*}[t!]
\centering
\caption{\textbf{Deployment Risk and Validation Framework for FMs in Computational Pathology.} This framework categorizes key barriers across deployment dimensions, identifying specific risks, technical root causes, observable symptoms, and corresponding detection and mitigation strategies.}
\label{tab:deployment_risk}
\resizebox{\textwidth}{!}{%
\begin{tabular}{@{}llllll@{}}
\toprule
\textbf{Deployment Dimension} & \textbf{Specific Risk} & \textbf{Technical Root Cause} & \textbf{Observable Symptom} & \textbf{Detection Strategy} & \textbf{Mitigation Strategy} \\ \midrule
Scanner variability & Feature distribution drift & Different optical pipelines & Performance drop across sites & Embedding distribution shift tests & Multi-scanner pretraining \\
Stain variability & Color space instability & Lab staining heterogeneity & False morphologic clustering & Stain-normalization stress tests & Stain-robust pretraining \\
Tissue prep artifacts & Non-biologic signal capture & Fold, chatter, necrosis & False high-confidence outputs & Artifact challenge sets & Artifact-aware augmentation \\
Label noise & Weak supervision contamination & Slide labels $\neq$ patch truth & Patch prediction inconsistency & Patch-slide disagreement metrics & Multi-instance training \\
Class imbalance & Embedding collapse & Rare class underrepresentation & Minority misclassification & Per-class calibration curves & Oversampling + metric learning \\
Population bias & Morphologic phenotype skew & Training cohort homogeneity & Site-specific overfit & Cross-ethnicity validation & Multi-population cohorts \\
Task overreach & Unsupported inference claims & Correlation $\neq$ causation & Molecular prediction overclaim & Mechanistic plausibility checks & Limit claim scope \\
Interpretability gap & Black-box embeddings & Latent space opacity & Reviewer skepticism & Saliency + concept activation & Hybrid symbolic overlays \\
Clinical workflow mismatch & Output not decision-aligned & Model-task mismatch & Low adoption & Pathologist usability studies & Task-aligned outputs \\
Regulatory barrier & Evidence insufficiency & Retrospective-only validation & Approval delay & Trial-grade evaluation & Prospective trials \\ \bottomrule
\end{tabular}%
}
\end{table*}

\section*{3.3 Safety, Liability, and Risk Management in Clinical AI Deployment}
\vspace{-1em}

As pathology AI systems expand beyond discriminative tasks such as classification toward generative applications including report drafting and virtual staining, the clinical risk profile shifts substantially. Whereas earlier systems primarily introduced risks of missed or delayed diagnoses, generative models introduce the possibility of producing plausible but factually incorrect content, commonly referred to as hallucinations\cite{rabbani2025generative,huang2025survey}. In pathology, this risk manifests in both textual and visual domains. Large language models(LLMs) may generate diagnostic reports citing nonexistent guidelines or inaccurate clinical context, while generative imaging models may produce synthetic structures, altered nuclear morphology, or distorted tissue architecture that mimic valid histological patterns\cite{kim2025medical}.  Because such artifacts often conform to the visual and linguistic logic of authentic pathology, they can be difficult to detect through routine review.

\noindent These technical limitations may reshape human decision making in clinical environments. Recent studies have documented automation bias, whereby clinicians may accept algorithmic recommendations over their own judgment, particularly under time pressure or cognitive load\cite{zajkac2023clinician}. This tendency is especially pronounced among trainees and early career practitioners, who may lack the experience required to critically interrogate confident but erroneous AI outputs\cite{dratsch2023automation}. Prolonged reliance on automated triage and decision support systems may also contribute to diagnostic deskilling, reducing exposure to rare variants and subtle morphological patterns that are essential for maintaining expert level visual acuity\cite{gaube2021do}. Over time, such shifts may erode professional competence and professional resilience.

\noindent The convergence of generative uncertainty and human cognitive bias has important legal and institutional implications. As AI systems increasingly influence diagnostic reasoning, errors attributable to algorithmic design, training data, or deployment configuration challenge traditional frameworks of medical malpractice\cite{maliha2021artificial, price2024liability}. Historically, software developers were often insulated under the learned intermediary doctrine, which assigned ultimate responsibility to physicians\cite{price2024liability}. However, emerging regulatory and judicial developments, including the European Union’s AI Act\cite{eu_ai_act_2024}, signal a growing tendency to classify clinical AI systems as regulated products subject to defective design and failure-to-warn claims (\hyperlink{supp:B7}{Details in Extended Section B.7}). Under this paradigm, liability may extend beyond individual clinicians to vendors and healthcare institutions\cite{price2024governance, muralidharan2024scoping}. 

\noindent Consequently, these evolving risk profiles underscore that successful clinical translation requires more than demonstration of algorithmic accuracy. Sustainable deployment depends on comprehensive governance frameworks encompassing human–AI interaction design, continuous performance auditing, formal incident reporting mechanisms, and rigorous quality assurance pipelines for generative outputs\cite{feng2025not, El_Arab2025, gilbert2023algorithm}. Without such safeguards, the combined effects of technical uncertainty, cognitive bias, and legal exposure are likely to constrain institutional willingness to adopt advanced pathology AI systems at scale.

%% file: sections/4-conclusion.tex
\noindent While pathology FMs  are rapidly accelerating biomarker discovery and pharmaceutical research \cite{CHIEF, diao2024human}, the demands of routine clinical practice diverge sharply from those of exploratory science. Clinical utility relies less on novel hypothesis generation and more on reliability, interpretability, and workflow compatibility\cite{vasey2022reporting}. In that light, this concluding section moves beyond discovery applications to focus strictly on clinical translation: transforming predictive scores into decision support, establishing sustainable economic infrastructure, and evolving the pathologist-AI relationship from supervision to collaboration\cite{Briganti2020-rl}.
\vspace{-1em}
\section*{4.1 From Predictive Models to Biologically Informed Clinical Decision-Making}
\vspace{-1em}

Clinical translation of pathology FMs  requires them to operate effectively in the complex and heterogeneous conditions of routine practice, beyond merely reproducing human-level performance in controlled academic settings\cite{bilal2025foundation}. Although benchmark performance metrics have reached near–human levels in retrospective evaluations, a persistent gap remains between algorithmic scores and tangible clinical benefit\cite{owusu2023imbalanced}. To enhance workforce efficiency, AI tools must move beyond standalone classification outputs and integrate organically into diagnostic workflows, leveraging computational strengths in quantitative tasks such as precise tumor burden estimation and mitotic counting that challenge human visual limitations\cite{vanderlaak2021deep,dvijotham2023enhancing}. Therefore, clinical deployment should not be  one-time static endpoint but adaptive to incorporate expert feedback and updates to diagnostic standards, including revisions to the WHO Classification of Tumors\cite{hyman2017implementing,NASRALLAH2023}. This iterative process helps ensure that the deployed models are aligned with contemporary clinical consensus\cite{El_Arab2025, gilbert2023algorithm}. However, it also imposes substantial maintenance efforts, including model updating, retraining, and revalidation. Without sustained institutional investment to support these activities, systems are likely to suffer from performance drift and increasing misalignment with clinical needs\cite{feng2025not}.

\noindent \noindent While robust infrastructure ensures operational reliability, the transformative potential of FMs  lies in their ability to deepen diagnostic insight\cite{GigaPath, CHIEF}.Beyond simple classification, these models are beginning to anchor histological features in molecular ground truth\cite{li2026ai, liu2025high}.Through multi-modal data integration, systems can map morphological patterns directly to local immune microenvironments and pathway activation states. This paradigm is supported by emerging evidence from histology-anchored spatial multi-omics\cite{liu2025high} and virtual spatial proteomics\cite{li2026ai}, which demonstrate the feasibility of inferring high-dimensional molecular phenotypes from standard tissue slides.This molecular grounding offers a potential calibration framework that may help histopathology FMs  better distinguish biologically meaningful signals from staining variability, artifacts, and benign mimics\cite{diao2024human}. Its practical value will largely depend on continued progress in co-registration, data scale, and protocol standardization. 

\noindent This capacity to biologically contextualize morphology becomes clinically decisive when human visual assessment reaches its limit. For example, generalist architectures like UNI enable the prediction of genomic biomarkers, such as MSI, directly from H\&E slides without task-specific retraining\cite{UNI}. In parallel, large-scale systems like Virchow provide granular prognostic stratification for rare and poorly differentiated malignancies, identifying high-risk phenotypes that escape conventional grading systems\cite{Virchow}. These capabilities position the standard H\&E slide as a scalable surrogate for molecular profiling, expanding the reach of precision oncology

\vspace{-1em}

\section*{4.2 From Research Prototypes to Sustainable Clinical and Economic Infrastructure}
\vspace{-1em}

A sustainable financial model is another core requirement for the successful translation of pathology FMs  into routine clinical practice\cite{homeyer2022artificial}. In many healthcare systems, AI diagnostic tools are not widely recognized as billable clinical services, which encourages healthcare providers to absorb the costs of development and deployment in order to generate revenue through downstream clinical activities. The United States operates within a decentralized reimbursement landscape in which pathology departments often treat AI systems as operational expenses\cite{ gondi2023paying}. China has a different government-led centralized pricing mechanism that makes qualified AI-assisted diagnostic services reimbursable\cite{evans2022regulatory, nhsa2025guide}. Healthcare providers in these systems differ in how they approach long-term investment, financial risk management, and resource commitments for maintaining and scaling clinical AI platforms\cite{}. Establishing a viable long-term ecosystem will therefore require sustainable business models that do not rely solely on traditional fee-for-service reimbursement\cite{stern2023economics}. Instead, greater emphasis on value-based care and cost effectiveness may be needed, in which AI is justified by improving diagnostic efficiency and accuracy, reducing clinical workload, and lowering downstream healthcare costs\cite{madabhushi2019artificial}. Advancing this model should be viewed as a long-term research and system-building effort, requiring coordinated support from public agencies, foundations, and healthcare institutions to develop hybrid funding and evaluation frameworks. 

\noindent Beyond macroeconomic policy, sustaining clinical AI systems requires a fundamental shift from resource intensive academic research practices to cost conscious, budget constrained clinical operations\cite{stern2023economics}. This transition necessitates rigorous economic evaluations to ensure that algorithmic utility justifies the recurring computational and operational expenditures. Most FMs  are developed in academic or industrial environments that rely on large computing clusters, extensive storage, and dedicated engineering teams that operate separately from day-to-day clinical workflows\cite{bommasani2021opportunities, Briganti2020-rl}.In real world clinical settings, FMs  are rarely plug and play solutions. Stringent data sovereignty frameworks, security protocols, and local governance policies often restrict deployment architectures, preventing the centralized data processing assumed in many academic workflows\cite{price2024governance}. Even when institutions are willing to adopt external models, direct off the shelf utilization is frequently compromised by substantial distributional shifts. Unlike the standardized interoperability of Health Information Systems, variations in scanner optics, staining protocols, and patient demographics introduce technical drifts that undermine generalizability\cite{lin2024federated}.  Consequently, successful deployment demands active localization, entailing rigorous on site validation, post hoc calibration, and, where necessary, site specific fine tuning\cite{Yang2023}.

\noindent Maintaining petabyte-scale storage and high-performance computing infrastructure creates substantial and recurring operational costs\cite{mulliqi2025foundation} that most hospital systems are unlikely to afford over time. We expect that more resource-efficient AI models operable on constrained hardware, paired with automated data collection and retraining to prioritize the most informative cases, will therefore be necessary. Emphasizing high-value datasets, such as rare cases, over indiscriminate data accumulation can help limit noise propagation while reducing the costs of manual curation and long-term storage\cite{altucci2025artificial}. A desired ecosystem suggests closer collaboration beyond a simple vendor–client relationship. This can support real-world feedback and ongoing model maintenance. However, it also raises practical questions about cost, staffing, and who is responsible for supporting highly trained technical personnel in clinical settings.

\vspace{-1em}

\section*{4.3 From Foundation Models to Clinical AI Agents: Human–AI Co-evolution}
\vspace{-1em}

Computational pathology has progressed from task-specific deep learning models to FMs , and is now beginning to move toward clinical AI agents. Over the past decade, vision-only deep learning models improved specific tasks such as tumor detection, cell segmentation, and grade classification\cite{litjens2017survey, Camelyon16, CoNSeP}. The subsequent emergence of visual FMs  and large language models marked a paradigm shift toward general-purpose frameworks, unifying multi-task execution with semantic, language-driven morphological interpretation~\cite{liu2025foundational,mehandru2024evaluating,teo2025generative}. Building on FMs , emerging clinical AI agents extend beyond multimodal prediction to support multi-step reasoning and task orchestration~\cite{chen2025pathagent,chen2025evidence,ghezloo2025pathfinder,sun2025cpathagent,wang2025pathology,ferber2025development,tu2025towards}. These systems combine vision models with large language models and external tools to retrieve prior knowledge, query clinical databases, integrate genomic and imaging data, and iteratively refine outputs in response to intermediate results. Rather than producing a single prediction, agents are designed to plan and execute sequences of actions, such as gathering missing information, consulting updated guidelines, and synthesizing evidence across sources\cite{sapkota2025ai, truhn2026artificial}. In this framework, agents optimize tasks to support end-to-end clinical reasoning. Their value lies in coordinating and amplifying existing diagnostic models and enabling diverse inputs and analytical requests in ways that more closely resemble how clinicians develop differential diagnoses and treatment plans\cite{rabbani2025generative, CONCH}.

\noindent There is growing agreement across both academia and industry that agent-based systems have the potential to reshape diagnostic workflows by shifting the human–AI relationship from supervision toward collaboration\cite{moor2023foundation}. In other domains, such systems have demonstrated how coordinated workflows can improve efficiency by distributing tasks across specialized models and tools. However, it remains challenging for such success to be directly mirrored in AI-based clinical diagnosis\cite{wang2025medagent}. Many component models in pathology are still not fully reliable, regardless of how often they are described as “generalizable.” These limitations can lead to error propagation in integrated agentic systems, in which small inaccuracies accumulate and amplify across stages\cite{altucci2025artificial,xu2025discovering}. Therefore, clinical AI agents will require additional effort in independent model validation and governance\cite{truhn2026artificial, feng2025not,alaa2025position}, to ensure diagnostic reliability and prevent clinical adoption from being driven by exaggeration. Ultimately, integrating agents into clinical practice should serve to extend the pathologist’s capabilities, allowing computational tools to handle data-intensive reasoning while ensuring final diagnoses remain grounded in expert oversight.

\vspace{-2em}
\begin{figure*}
\setstretch{0.9}
  \centering
  \includegraphics[width=\linewidth]{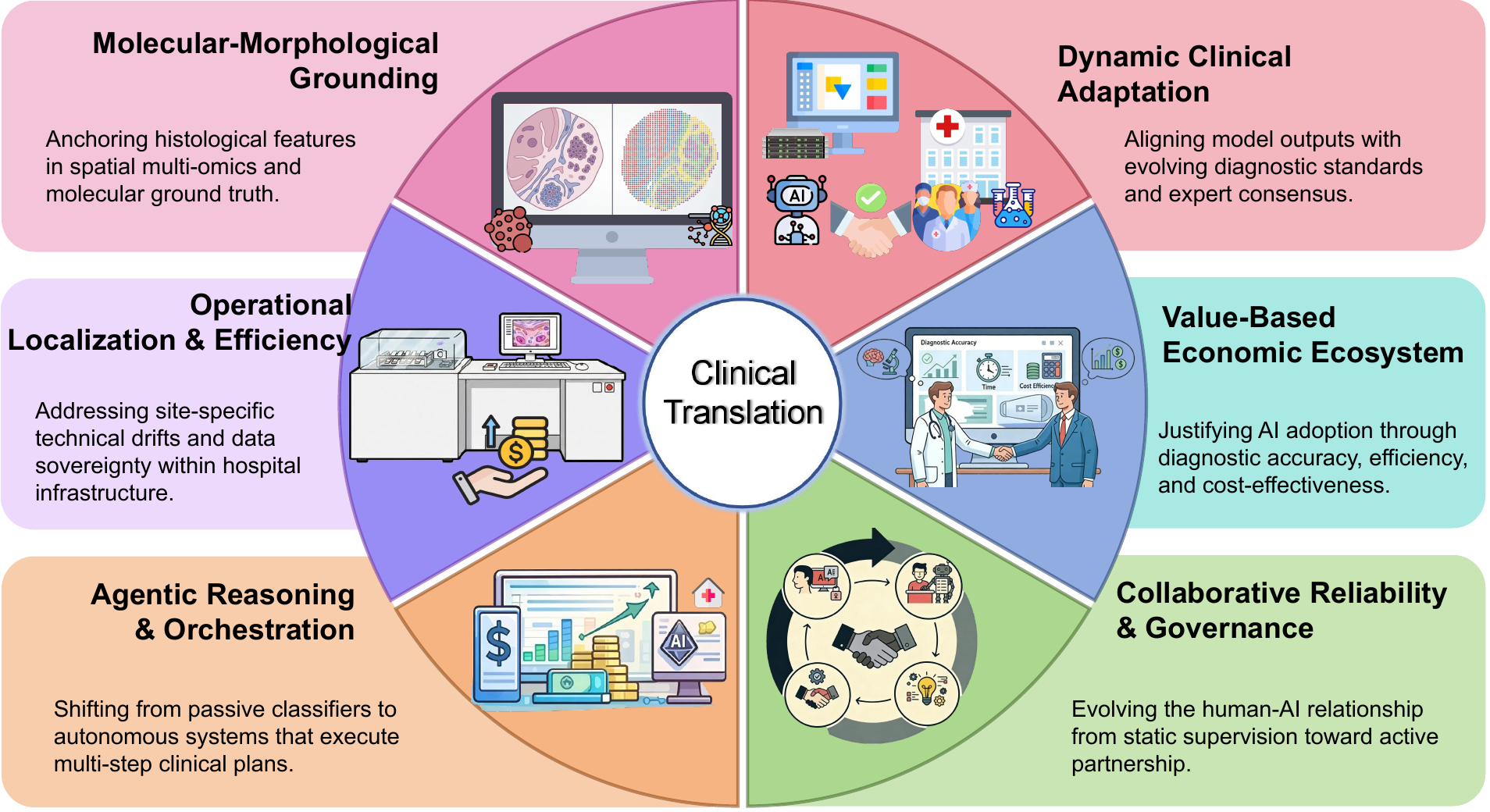}
        \caption{\textbf{Major Challenges and Translational Opportunities for Clinical Implementation of Pathology FMs } We outline Six key directions for advancing the clinical translation of pathology AI FMs . (1) Bridge the gap between technical capabilities and clinical priorities to address subtle, context-dependent diagnostic needs. (2) Improve robustness to pre-analytic variability and inter-institutional heterogeneity to enhance generalizability. (3) Harness generative models for applications like virtual staining and automated reporting, while managing risks related to safety, misinformation, and accountability. (4) Develop multimodal systems that integrate histology with transcriptomics, imaging, and clinical metadata for deeper, personalized disease profiling. (5) Build scalable, cost-efficient infrastructure to support model updates and high-resolution image processing. (6) Navigate evolving ethical, legal, and regulatory frameworks to ensure diagnostic reliability and mitigate the risks of error propagation. }
\label{fig:opportunities}
\end{figure*}

%% file: sections/5-perspective.tex
\noindent Computational pathology is entering a new era in which generalizable and integrative AI systems are transitioning from retrospective analytical tools to prospective components of clinical infrastructure. Advances in foundation and agentic models now enable multimodal diagnostic support, virtual molecular assays, and workflow-integrated decision assistance, yet real-world adoption remains constrained by economic, technical, and governance barriers. Realizing clinical value will therefore require shifting emphasis from predictive accuracy toward reliability, interoperability, economic sustainability, and institutional accountability within routine diagnostic workflows. Future success will depend on system-level alignment with regulatory, institutional, and reimbursement frameworks that enable scalable and accountable clinical deployment across diverse practice environments.

%% file: sections/5-appendix.tex
\newpage
\begingroup
\setlength{\parskip}{0.2in}
\setlength{\parindent}{0pt}

\hypertarget{supp:A}{\Heading{Extended Section A: Specialization and Generalization in Computational Pathology}}
\vspace{0.3em}

This section provides the technical background underlying the foundation models discussed in the main text. It covers the architectural paradigm shift from task-specific to foundation models and their clinical trade-offs (A.1), the self-supervised learning paradigm that enables data-efficient training (A.2), a reference catalog of clinical applications (A.3), and a glossary of key technical concepts referenced throughout the main text (A.4).

\hypertarget{supp:A1}{\heading{A.1 Encoder Architectures and Their Clinical Trade-offs}}

The choice of encoder architecture has direct implications for clinical deployment, extending beyond benchmark performance to encompass interpretability, computational feasibility, and regulatory considerations.

\hypertarget{supp:A1:paradigm}{\hheading{From Task-Specific Models to the Foundation Model Paradigm}}

The evolution from task-specific models (TSMs) to foundation models (FMs) represents a fundamental paradigm shift in computational pathology, rather than a simple replacement of one technology by another. As discussed in Main Text Section 2, TSMs continue to offer competitive performance with significantly lower computational overhead when labeled data is abundant and clinical tasks are narrowly defined. The architectural progression from convolutional neural networks (CNNs)\cite{lecun1989backpropagation} to Vision Transformers (ViTs)\cite{ViT} provided the technical substrate upon which this shift was built, enabling the self-supervised, data-efficient training strategies that define modern foundation models.

CNNs operate through hierarchical feature extraction via fixed-size convolutional kernels, capturing local patterns through progressively expanding receptive fields\cite{srinidhi2021deep}. This inductive bias makes CNNs effective for detecting local morphological features, and indeed CNN-based TSMs remain the workhorses of many deployed clinical applications, including mitotic figure detection and nuclear segmentation\cite{wang2023retccl,aubreville2023mitosis,CoNSeP}. However, their inherent limitation in modeling explicit long-range spatial dependencies constrains their capacity to capture the complex, multi-scale tissue relationships that characterize tumor microenvironments and that are essential for the higher-order diagnostic tasks discussed in Main Text Sections 2.2–2.4.

ViTs address this limitation through self-attention mechanisms that enable global context modeling from the first layer, allowing dynamic weighting of all image patches regardless of spatial distance\cite{xu2023vision}. This architectural property has proven advantageous for capturing the complex spatial relationships characteristic of tumor microenvironments\cite{li2023vision}. Critically, the self-attention mechanism also enables ViTs to learn rich, transferable representations from large, uncurated, unlabeled datasets through self-supervised learning (SSL): the paradigm shift that distinguishes foundation models from their task-specific predecessors (detailed in Section A.2). By removing the dependency on expert-annotated labels, the ViT+SSL combination allowed training on datasets of unprecedented scale (e.g., UNI\cite{UNI} on 100+ million patches, and Virchow\cite{Virchow} on 2 billion patches), producing representations that generalize across the diverse downstream tasks cataloged in Main Text Section 2.1. However, this capability comes with quadratic computational complexity with respect to input size, presenting practical constraints for gigapixel whole-slide image processing.

Hybrid architectures have emerged to bridge these paradigms: CTransPath\cite{wang2022Ctranspath} combines a CNN backbone with a multi-scale Swin Transformer module, leveraging semantically-relevant contrastive learning to capture both local texture patterns and broader contextual features while maintaining computational tractability. Such hybrids illustrate that the TSM–FM boundary is not absolute. Rather, the field has evolved along a continuum where architectural innovations progressively expanded the scope of learnable representations from narrow, task-specific features to the generalizable, multi-scale embeddings that underpin the clinical applications detailed in the main text\cite{springenberg2023modern,chen2022scaling,filiot2023scaling}.

From an interpretability standpoint, the architectures demand different explanation approaches. CNNs require post-hoc attribution methods such as Gradient-weighted Class Activation Mapping (Grad-CAM) to generate visual explanations, which compute importance scores by backpropagating gradients to identify discriminative regions\cite{dorrich2023explainable,holscher2025decoding}. While effective, these methods provide approximations of model reasoning rather than direct insight into decision processes. ViTs offer potential advantages through their inherent attention mechanisms, which provide native interpretability by explicitly encoding patch-to-patch relationships\cite{li2023vision}. Attention maps can be directly extracted to visualize which tissue regions the model considered most relevant, though recent comparative studies indicate that naive attention visualization can be prone to artifacts, and more sophisticated methods such as ViT-Shapley may be required for reliable clinical explanations\cite{supp_a1_vitshapley2025}.

The convergence of ViT architectures with self-supervised pretraining has unlocked a cascade of technical capabilities that were infeasible under the TSM paradigm, directly enabling the clinical applications discussed across Main Text Section 2. First, the rich, generalizable representations learned through SSL enable zero-shot and few-shot classification (Section A.4), wherein models such as CONCH\cite{CONCH} and PLIP\cite{PLIP} classify previously unseen histological entities without task-specific training, a capability central to the rare disease retrieval discussed in Main Text Section 2.1. Second, vision-language alignment models (e.g., CONCH\cite{CONCH}, MUSK\cite{MUSK}) jointly embed histological images and diagnostic text into a shared semantic space, enabling natural language-guided case retrieval and the automated report generation described in Main Text Section 2.4.2. Third, conversational diagnostic assistants such as PathChat\cite{PathChat} leverage these multimodal representations to support interactive clinical reasoning, incorporating patient history and domain-specific terminology. Fourth, the emerging agentic AI frameworks (e.g., SlideSeek\cite{chen2025evidence}), discussed in Main Text Section 2.4.3, exploit hierarchical “Supervisor-Explorer” architectures that autonomously navigate gigapixel slides to gather diagnostic evidence, a capability predicated on the robust, multi-scale representations that only foundation-scale models provide. Each of these advances builds upon the same foundational shift: the ability to learn expressive representations at scale without manual annotation.

For clinical deployment, these interpretability differences carry both regulatory and practical implications. Explainability requirements under frameworks such as the EU AI Act and FDA guidance increasingly demand that clinical AI systems provide justification for their outputs (see Extended Section B.7)\cite{supp_a1_busch2024euaiact,supp_a1_minssen2024euaiact_mdp}. Architectures offering more direct interpretability pathways may face fewer barriers to regulatory approval, though the quality and clinical relevance of explanations remains the critical determinant\cite{holscher2025decoding}. Importantly, interpretable foundation model outputs also offer a potential solution to the longstanding inter-observer variability problem in pathology (Main Text Section 2). When model attention maps and feature attributions correlate with established histopathological knowledge, they can serve as a standardizing reference that reduces subjective diagnostic variation across pathologists. Conversely, when models highlight previously unrecognized morphological patterns associated with clinical outcomes (as demonstrated by survival prediction models identifying sub-visual stromal features in the Main Text Section 2.3)interpretable outputs can catalyze the discovery of novel histopathological biomarkers, further bridging the gap between computational and clinical pathology.

\hypertarget{supp:A1:patchVSencod}{\hheading{Patch-level vs.\ Slide-level Encoders}}

Foundation models can be categorized by their operational granularity: patch-level encoders process fixed-size tiles (typically 224×224 or 256×256 pixels) independently, while slide-level encoders aim to capture whole-slide context directly. Patch-level encoders such as UNI\cite{UNI} and Virchow\cite{Virchow} generate rich local representations but require downstream aggregation, via attention-based pooling, transformers, or graph neural networks, to produce slide-level predictions. Slide-level encoders such as Prov-GigaPath\cite{GigaPath} and TITAN\cite{TITAN} attempt to model inter-patch relationships during encoding, potentially capturing tissue architecture more holistically. Vision-language foundation models such as CONCH\cite{CONCH}, pretrained on over 1.17 million pathology image-caption pairs, represent a third paradigm that aligns visual and textual representations, enabling zero-shot transfer and natural language-guided retrieval across diverse histopathologic tasks. For clinical workflows, this distinction influences both inference latency and the types of clinical questions addressable: patch-level systems excel at localized tasks (e.g., mitotic detection) while slide-level systems may better support global assessments (e.g., tumor staging), multimodal synthesis (Main Text Section 2.4), and the virtual spatial biology applications described in Main Text Section 2.2.2.

\hypertarget{supp:A1:frozen}{\hheading{The Frozen Encoder Paradigm}}

A critical deployment consideration is the frozen encoder paradigm, wherein large pretrained encoders are fixed during downstream task adaptation, with only lightweight classification heads trained on task-specific data\cite{CONCH}. This approach offers several practical advantages: it dramatically reduces computational requirements for adaptation, enables deployment on more modest hardware, and creates a clear separation between the validated foundation model and task-specific components. This separation has regulatory implications, as it allows institutions to maintain a stable, versioned encoder while updating downstream classifiers without full revalidation\cite{abels2019computational}. Recent clinical deployments have demonstrated that frozen encoder approaches combined with simple aggregators can achieve clinical-grade performance across multiple tasks\cite{Campanella2025ClinicalBenchmark}, though the trade-off between frozen and fine-tuned approaches remains task-dependent, with full fine-tuning sometimes providing meaningful performance gains when sufficient computational resources are available\cite{Campanella2025RealWorldDeployment}.
Critically, the frozen encoder paradigm enables the scalable automation of diagnostic workflows highlighted throughout Main Text Section 2. Because a single, validated encoder can serve as the backbone for multiple downstream tasks, institutions can deploy parallel clinical applications (tumor detection, histological grading, molecular biomarker prediction in Main Text Section 2.2.1, and survival stratification in Main Text Section 2.3) without retraining or revalidating the core model for each task. For example, the same UNI encoder has been adapted to support both rare cancer classification across 108 OncoTree categories and routine pan-cancer detection, while TITAN’s frozen representations support simultaneous survival prediction and zero-shot rare disease diagnosis\cite{TITAN}. This “one encoder, many heads” deployment pattern directly addresses the scalability challenge of translating foundation models into clinical laboratories operating across diverse case mixes.

\hypertarget{supp:A2}{\heading{A.2 Self-Supervised Learning: Why It Matters for Translation}}

Self-supervised learning (SSL) has emerged as the dominant paradigm for training pathology foundation models, fundamentally altering the economics and feasibility of developing clinically useful AI systems. Understanding the translational implications of SSL is essential for evaluating foundation model deployment.

\hypertarget{supp:A2:Bottleneck}{\hheading{Eliminating the Annotation Bottleneck}}

The central translational advantage of SSL is its elimination of the annotation bottleneck that has historically constrained supervised learning in pathology. Traditional supervised approaches require expert pathologists to annotate training examples, a process that is expensive, time-consuming, and fundamentally unscalable for the thousands of diagnostic entities encountered in clinical practice. Modern SSL encompasses a diverse family of pretraining strategies, each exploiting different forms of self-supervision to learn rich visual representations without manual labels. Contrastive learning methods (e.g., MoCo\cite{MOCO_2020}, SimCLR\cite{chen2020simple}) learn by maximizing agreement between augmented views of the same image while pushing apart representations of different images. Self-distillation approaches (e.g., DINO\cite{caron2021emerging}, DINOv2\cite{oquab2023dinov2} and DINOv3\cite{dinov3}) employ a teacher-student framework wherein a momentum-updated teacher network produces soft targets for the student, enabling the learning of semantically rich, locally structured features particularly well-suited to dense prediction tasks in pathology. Masked image modeling (e.g., MAE\cite{he2022masked}, iBOT\cite{zhou2021ibot}) reconstructs randomly masked patches, forcing the model to learn contextual relationships between tissue regions. These strategies can also be combined: DINOv2 jointly optimizes a self-distillation objective with masked image modeling, and domain-specific approaches such as CTransPath\cite{wang2022Ctranspath} employ semantically-relevant contrastive learning on 15,000+ WSIs to produce pathology-tailored representations. Table~\ref{table:list of model} summarizes the various pretraining strategies used by recent pathology foundation models.

This diversity of SSL methods has enabled training on an unprecedented scale. Models like UNI\cite{UNI} (pretrained on over a hundred thousand whole slide images) and Virchow\cite{Virchow} (pretrained on 2 billion patches from 1.5 million slides) develop representations that transfer effectively across the diverse downstream tasks discussed in Main Text Section 2.

\hypertarget{supp:A2:RareDisease}{\hheading{Rare Disease and Long-Tail Coverage}}

Perhaps the most clinically significant implication of SSL is its potential to address rare and underrepresented diseases, entities for which supervised learning is inherently limited by data scarcity. This directly underpins the “long-tail” retrieval capabilities highlighted in Main Text Section 2.1. Pathology confronts a severe long-tail distribution: while common cancers may be represented by thousands of cases in institutional archives, rare subtypes may have only a handful of examples available for model development. Foundation models pretrained via SSL have demonstrated strong few-shot capabilities, enabling slide-level classification using only 1–4 annotated examples per class through prototype-based approaches\cite{TITAN}. UNI, for instance, excels specifically at classifying rare and underrepresented diseases, including 90 of 108 rare cancer types in OncoTree classification, suggesting that SSL pretraining captures generalizable histopathologic features that transfer to conditions rarely encountered in training data. Similarly, TITAN, a multimodal whole-slide foundation model, demonstrates capability for rare cancer diagnosis and survival prediction without requiring fine-tuning, specifically by leveraging vision-language alignment from pathology reports\cite{TITAN}.

\hypertarget{supp:A2:DomainAdapt}{\hheading{Domain Adaptation and Pre-training Data Considerations}}

While SSL enables training without labels, the composition of pre-training data significantly influences downstream performance. Models pretrained on ImageNet (natural images) consistently underperform pathology-specific models, reflecting the substantial domain gap between everyday objects and histological tissue\cite{Campanella2025ClinicalBenchmark}. However, evidence suggests that for disease detection tasks, models trained with DINO or DINOv2 achieve comparable performance regardless of differences in pretraining dataset size or composition, whereas biomarker prediction tasks show greater sensitivity to pretraining data characteristics. Tellez, Litjens, and colleagues\cite{tellez2019quantifying} demonstrated that stain color augmentation during training can substantially reduce the domain gap across institutions, with data augmentation strategies tailored to H\&E-stained histopathology (including color perturbation, morphological transformations, and stain-specific augmentation)improving cross-dataset generalization by up to 10 percentage points on external validation\cite{faryna2021tailoring}\cite{faryna2024automatic}. This distinction has practical implications: institutions developing task-specific models must consider whether their target application resembles detection (where general pathology features suffice) or requires more specialized representations.

The choice of SSL algorithm also matters. Comparative studies indicate that DINOv2 tends to outperform earlier contrastive learning approaches (such as MoCo) for pathology pretraining, though DINO and DINOv2 achieve similar downstream performance despite DINOv2’s greater computational cost\cite{Campanella2025ClinicalBenchmark}. For institutions with limited computational resources, this suggests that well-executed training with DINO on carefully curated pathology data may achieve results comparable to more expensive DINOv2 training.

\hypertarget{supp:A2:BiasandGeneralization}{\hheading{Implications for Bias and Generalization}}

SSL models inherit biases from their pretraining data, raising considerations for clinical deployment that directly connect to the domain shift and generalization challenges detailed in Main Text Section 3.2 and Extended Section B.5\cite{bommasani2021opportunities}\cite{Zhao2025-ur}. Models pretrained predominantly on data from academic medical centers in high-income countries may encode systematic biases related to patient demographics, staining protocols, and scanner characteristics. While the field has not yet systematically characterized these biases, institutions deploying foundation models should consider the alignment between their patient population and the model’s training distribution. The provenance of pretraining data (including geographic diversity, tissue type representation, and staining protocol coverage) represents an underexplored dimension of model selection that may prove critical for equitable deployment across diverse clinical settings.

\hypertarget{supp:A3}{\heading{A.3 Clinical Applications Reference}}

Basic applications primarily automate existing diagnostic workflows. Tumor detection models identify malignant regions within whole-slide images, thereby facilitating case triage and quality control. Grading systems further quantify key histological features, such as Gleason patterns in prostate cancer and mitotic figures in breast cancer, often achieving reproducibility that exceeds inter-observer agreement among pathologists. In addition, subtype classification models distinguish morphologically similar entities, such as lung adenocarcinoma and squamous cell carcinoma, providing critical support for treatment selection\cite{echle2021deep, ochi2025pathology}.

Advanced applications extract information beyond conventional microscopy capabilities: molecular biomarker prediction infers genomic alterations (e.g., microsatellite instability, BRAF mutations, IDH status) directly from H\&E morphology, potentially reducing turnaround time and cost compared to molecular testing\cite{zimmermann2024virchow2}. Survival prediction models further stratify patients into prognostic groups by leveraging morphological patterns that are not readily discernible to human observers. Validated systems have reported hazard ratios comparable to or even exceeding those of established clinical staging frameworks\cite{CHIEF}. In addition, treatment response prediction supports end-to-end inference of therapeutic outcomes from pre-treatment histology. Representative applications include predicting immunotherapy response in melanoma and non-small cell lung cancer\cite{johannet2021predicting, rakaee2025deepio}. The MUSK\cite{MUSK} vision-language model, trained on over 50 million pathology images and 1 billion pathology text tokens, extends this capability through multimodal integration, demonstrating state-of-the-art performance in melanoma relapse prediction and immunotherapy response assessment across diverse cancer types. Wang et al. further demonstrated foundation model-based prognosis prediction for gastrointestinal cancers, with models identifying patients most likely to benefit from adjuvant therapy\cite{wang2025foundation}. Complementary spatial feature approaches, such as Corredor, Madabhushi, and colleagues’ work on collagen disorder architecture (CoDA)\cite{li2021collagen} and spatial immune cell organization, provide interpretable morphological biomarkers that capture microenvironment characteristics relevant to prognosis and treatment response in colon and lung cancers\cite{corredor2019spatial}\cite{nag2025coda}.

Foundation models excel particularly in data-limited scenarios: rare cancer detection benefits from transfer learning, with models like UNI and Virchow demonstrating strong few-shot performance across uncommon tumor types where dedicated training datasets are unavailable\cite{zimmermann2024virchow2}. 

\hypertarget{supp:A4}{\heading{A.4 Key Technical Concepts Referenced in the Main Text}}

\textbf{Zero-shot classification} refers to the ability of a model to classify entities from categories not explicitly represented in its training data. In pathology foundation models, this is typically achieved through vision-language alignment: models such as CONCH\cite{CONCH} and PLIP\cite{PLIP} learn a shared embedding space between histological images and textual descriptions during pretraining, enabling classification of novel tissue types by comparing image embeddings against text-based category descriptors (e.g., “papillary thyroid carcinoma”) without any task-specific labeled examples. This capability directly enables the rare disease retrieval applications discussed in Main Text Section 2.1, where pathologists can query digital archives using diagnostic language rather than curated training sets.

\textbf{Few-shot classification} extends this to settings where a small number of labeled examples (typically 1–16 per class) are available. Foundation models leverage their pretrained representations to generalize from these minimal exemplars through prototype-based methods, nearest-neighbor retrieval, or lightweight linear probes. This capability is particularly valuable for rare tumor subtypes where comprehensive labeled datasets are infeasible.

\textbf{Multi-instance learning (MIL)} is the dominant paradigm for whole-slide image analysis in computational pathology\cite{dietterich1997solving}. Because WSIs are too large to process as single inputs, they are divided into thousands of smaller patches (instances), which are collectively treated as a “bag.” In the standard MIL formulation, a bag-level label (e.g., tumor-positive) is assigned to the entire slide, while individual patch labels remain unknown. Attention-based MIL aggregators (e.g., ABMIL\cite{ilse2018attention}) learn to weight patches by diagnostic relevance, enabling both slide-level predictions and interpretable attention heatmaps that highlight diagnostically important regions. This framework underpins the slide-level analyses discussed throughout Main Text Sections 2.2–2.4, including the generative reporting models (Main Text Section 2.4.2) where decoder architectures dynamically query patch-level features via cross-attention.

\textbf{Vision-language pretraining} jointly trains models on paired image and text data to learn aligned multimodal representations. In pathology, this involves training on histological image–diagnostic text pairs drawn from pathology reports, captions, or curated datasets such as PathAlign\cite{PathAlign} and PathGen\cite{PathAlign}. The resulting models can perform cross-modal retrieval (finding images matching a text query, or vice versa), zero-shot classification, and automated report generation. This paradigm underlies the multimodal synthesis capabilities discussed in Main Text Section 2.4 and the CBIR systems of Section 2.1.

\hypertarget{supp:B}{\Heading{Extended Section B: Challenges to Translation: Economic and Technical Barriers}}
\vspace{0.3em}

This section provides detailed evidence on the technical, validation, infrastructure, and regulatory barriers that impede translation of computational pathology models into routine clinical use. As the main text argues, the disconnect between academic success and clinical adoption reflects a deeper structural misalignment between research incentive structures and the practical requirements of sustained deployment\cite{homeyer2022artificial} \cite{vanderlaak2021deep} . The subsections below are organized to follow and extend the structure of the main text: computational and infrastructure constraints (B.1–B.3, supporting Main Text Section 3.1), the benchmark-to-clinic reality gap and domain shift (B.4–B.6, supporting Main Text Section 3.2), and regulatory considerations (B.7, supporting Main Text Section 3.3).

\hypertarget{supp:B1}{\heading{B.1 Computational Realities in Hospital Settings}}

The translation of foundation models from research benchmarks to clinical deployment hinges on computational feasibility within real-world hospital infrastructure. While academic publications typically report performance on high-end GPU clusters, clinical laboratories operate under fundamentally different constraints that directly impact model selection and deployment architecture. These computational realities represent a key component of the cost–benefit asymmetry discussed in Main Text Section 3.1: the financial burden of deploying AI falls disproportionately on institutions that must procure, maintain, and operate the inference infrastructure.

\hypertarget{supp:B1:InfLatency}{\hheading{Inference Latency and Clinical Workflow Integration}}

Clinical pathology operates under strict turnaround time requirements that vary by use case. Intraoperative consultations, such as frozen section evaluation, demand results within minutes, while routine diagnostic workflows may tolerate longer processing times but still require predictable throughput to avoid bottlenecks\cite{rad2025deep}. A typical diagnostic whole-slide image containing 50,000–100,000 patches requires billions of attention computations when processed through transformer-based architectures, translating to 15–45 minutes of processing time on standard clinical hardware, a latency fundamentally incompatible with workflows requiring sub-5-minute turnaround\cite{rad2025deep}. This computational reality has driven development of sparse attention mechanisms and hierarchical processing strategies, which are not merely algorithmic optimizations but clinical necessities for real-world deployment.

\hypertarget{supp:B1:Memory}{\hheading{Memory Footprints and Hardware Requirements}}

Foundation models vary substantially in their computational demands. Recent benchmarking demonstrates that GPU memory requirements range from approximately 0.13 GB for smaller models (e.g., ViT-Small with 22M parameters) to over 4.6 GB for larger architectures such as H-optimus-0, with widely deployed models like UNI requiring approximately 1.2 GB and Virchow2 requiring approximately 4.6 GB\cite{Campanella2025ClinicalBenchmark}. These differences have direct implications for the capital investment burden described in Main Text Section 3.1: while research institutions may access NVIDIA A100 clusters, typical hospital IT infrastructure often comprises more modest hardware. Running models like Prov-GigaPath requires high-end GPUs that represent substantial capital investment for resource-constrained institutions\cite{Ma2025GPFM}\cite{Filiot2025H0mini}. Notably, smaller models trained with appropriate self-supervised learning algorithms can achieve comparable performance to larger counterparts on detection tasks, suggesting that model selection should balance performance requirements against institutional hardware realities\cite{grashei2025pathryoshka}.

\hypertarget{supp:B1:PracticalEfficency}{\hheading{Practical Efficiency Considerations}}

Inference throughput, measured in tiles processed per second, varies considerably across foundation models and directly impacts clinical utility. Models offering favorable performance-to-compute ratios(such as ViT-Small architectures achieving comparable detection accuracy to larger models) may be preferable for high-volume clinical laboratories where throughput matters as much as peak accuracy. The development of smaller, distilled models that retain much of their larger counterparts’ performance represents a promising direction for democratizing access to foundation model capabilities\cite{Ma2025GPFM}. Recent work has demonstrated that models requiring only 35 million parameters can train on consumer-grade workstations (single NVIDIA RTX 4090), substantially lowering the barrier to entry compared to billion-parameter models requiring datacenter-scale infrastructure\cite{OpenSlideFM2025}.

\hypertarget{supp:B2}{\heading{B.2 Integration with Laboratory Information Systems}}

The interoperability challenges described here represent a concrete manifestation of the “fragmented translational landscape” identified in the main text introduction, where digital pathology’s lack of standardization, unlike radiology’s DICOM-based workflows, creates systemic barriers to AI integration.

\hypertarget{supp:B2:Interoperability}{\hheading{Interoperability Standards}}

Pathology AI deployment requires bidirectional communication between AI systems and existing laboratory infrastructure, primarily the Laboratory Information System (LIS), Picture Archiving and Communication System (PACS), and Electronic Health Record (EHR). The Digital Imaging and Communications in Medicine (DICOM) standard, extended through Supplement 145 for whole-slide imaging, provides the foundational framework for image data exchange\cite{Herrmann2018DICOM}. However, mapping proprietary LIS data structures to standard HL7 specimen representations remains nontrivial and frequently requires vendor-specific customization or third-party middleware\cite{Herrmann2018DICOM}.

The Integrating the Healthcare Enterprise (IHE) Pathology and Laboratory Medicine (PaLM) domain has established workflow profiles for digital pathology interoperability, including the Digital Pathology Image Acquisition (DPIA) profile that defines communication between scanners, image management systems, and archives\cite{dash2021integrating}. Despite these standards, real-world implementations often encounter gaps between the HL7 specimen data model and DICOM requirements, necessitating custom integration work.

\hypertarget{supp:B2:Proprietary}{\hheading{Proprietary Format Challenges}}

Whole-slide scanners generate images in vendor-specific formats (e.g., Aperio .svs, Hamamatsu .ndpi, Philips .tiff, Leica .scn), creating interoperability barriers\cite{dash2021integrating}. As noted in Main Text Section 3.2, this fragmentation contrasts sharply with radiology’s standardized DICOM workflows and constitutes a fundamental deployment bottleneck. Converting these formats to DICOM Supplement 145-compliant objects enables integration with enterprise PACS and vendor-neutral archives (VNAs), but requires additional processing infrastructure\cite{clunie2021dicom}. Many institutions deploy middleware solutions that ingest proprietary formats, perform standards-based conversion, enrich DICOM headers with LIS/EHR metadata, and generate HL7 Instance Availability Notifications for downstream systems\cite{clunie2021dicom}.

The absence of standardized notification protocols between scanners and LIS/PACS platforms complicates real-time workflow integration\cite{ClosingGap2025}. AI systems requiring immediate access to newly scanned slides must interface with notification mechanisms that many legacy systems do not support natively.

\hypertarget{supp:B3}{\heading{B.3 Deployment Topology Options}}

The choice of deployment topology directly shapes the cost structure and risk distribution discussed in Main Text Section 3.1\cite{00aggarwal2025artificial}. Each option entails distinct trade-offs between capital expenditure, operational overhead, data sovereignty, and scalability.

\hypertarget{supp:B3:Onpremise}{\hheading{On-Premises Deployment}}

On-premises deployment maintains complete data sovereignty, with all patient images and AI inference occurring within institutional infrastructure. This approach addresses data residency requirements and eliminates concerns about protected health information (PHI) transmission across institutional boundaries\cite{Zarella2023DeploymentConsiderations}\cite{yadav2023data}. However, it requires substantial capital investment in GPU hardware capable of running foundation models (often NVIDIA A100 or similar high-end accelerators) for billion-parameter models\cite{Campanella2025RealWorldDeployment}.

Institutions pursuing on-premises deployment must maintain IT expertise for system administration, model updates, and performance monitoring. The total cost of ownership includes not only hardware acquisition but ongoing maintenance, power, cooling, and eventual hardware refresh cycles, contributing to the “persistent operational liability” described in Main Text Section 3.1.

\hypertarget{supp:B3:Cloud}{\hheading{Cloud-Based Deployment}}

Cloud platforms offer scalable GPU resources without capital expenditure, enabling institutions to deploy foundation models using consumption-based pricing. Major cloud providers (AWS, Azure, Google Cloud) offer HIPAA-eligible services with Business Associate Agreements (BAAs) that establish legal frameworks for PHI handling\cite{FoleyLardner2025HIPAA}. Cloud deployment requires careful configuration: data must be encrypted at rest and in transit, access controls must enforce role-based permissions, and audit logs must capture all PHI access for compliance verification\cite{Ardon2024FinancialAspects}.

Cloud deployment introduces network dependencies: gigapixel whole-slide images (typically 1–3 GB each) must be transmitted to cloud infrastructure, creating bandwidth bottlenecks for high-volume laboratories\cite{Komura2024PathologyFoundationModels}. Latency-sensitive applications such as intraoperative consultation may be impractical with cloud-based inference unless edge preprocessing reduces data transmission requirements.

\hypertarget{supp:B3:Hybrid}{\hheading{Hybrid Architectures}}

Hybrid approaches combine on-premises preprocessing with cloud-based inference, balancing data sovereignty with computational scalability. Local systems perform tile extraction, quality control, and initial filtering, transmitting only relevant image regions to cloud infrastructure for foundation model inference\cite{Zarella2023DeploymentConsiderations}\cite{Zarella2023DeploymentConsiderations}. This reduces bandwidth requirements and limits PHI exposure while leveraging cloud GPU resources for computationally intensive tasks.
Containerized microservices (e.g., Docker-based deployments) facilitate consistent AI deployment across hybrid environments, enabling institutions to run identical inference pipelines on-premises or in cloud infrastructure with minimal configuration changes.

\hypertarget{supp:B4}{\heading{B.4 The Benchmark-Reality Gap}}

This section provides empirical evidence for the “persistent gap between technical assumptions and the realities of routine diagnostic practice” identified in Main Text Section 3.2. The limitations described here (confounded benchmarks and hidden stratification) illustrate why models achieving expert-level performance on curated datasets may fail under clinical conditions.

\hypertarget{supp:B4:Limitations}{\hheading{Limitations of Standard Datasets}}

TCGA has served as the de facto benchmark for computational pathology research, providing researchers access to thousands of whole-slide images with associated clinical and molecular annotations. However, TCGA’s utility as a proxy for clinical performance is fundamentally limited. Deep neural networks trained on TCGA images can accurately classify the acquisition site even when trained for ostensibly unrelated tasks such as cancer type classification\cite{Dehkharghanian2023BiasedData}. This finding reveals that models may learn institution-specific artifacts (staining protocols, scanner characteristics, tissue processing variations) rather than diagnostically relevant histomorphologic features. When such models encounter slides from institutions not represented in training data, performance degrades unpredictably.

The multi-institutional composition of TCGA, while intended to promote diversity, paradoxically enables models to exploit site-specific confounders. Slides from different hospitals exhibit systematic differences in color intensity, tissue preparation, and scanning hardware that constitute learnable patterns orthogonal to pathologic diagnosis\cite{Howard2021SiteSpecificSignatures,Dehkharghanian2023BiasedData}. Models trained on TCGA data have been shown to falter when applied to slides from community hospitals, where staining protocols and scanner equipment differ from academic medical centers that predominate in research datasets\cite{00aggarwal2025artificial}. Tools such as HistoQC\cite{janowczyk2019histoqc}, developed by Janowczyk, Madabhushi, and colleagues, address this gap by providing automated, open-source quality control for histopathology images, detecting pen markings, tissue folds, out-of-focus regions, and staining artifacts that confound model performance and contribute to the benchmark-reality gap.

\hypertarget{supp:B4:Hidden}{\hheading{The Hidden Stratification Problem}}

Aggregate performance metrics can mask clinically meaningful failures on important patient subgroups, a phenomenon termed “hidden stratification”\cite{OakdenRayner2020HiddenStratification}. This directly compounds the diagnostic complexity discussed in Main Text Section 3.2, where clinical pathology rarely conforms to closed-set classification and individual slides frequently contain complex mixtures of malignant, benign, inflammatory, and reactive processes. A cancer detection model may achieve 95\% overall accuracy while consistently missing a rare but aggressive subtype, because standard evaluation schemas incompletely describe the meaningful variation within disease categories. A “lung cancer” label encompasses both solid and subsolid tumors, central and peripheral neoplasms, and dozens of histologic subtypes with distinct clinical implications. When these subgroups are not explicitly labeled in test sets, even held-out evaluation may provide falsely reassuring performance estimates.

Studies have demonstrated that hidden stratification can produce performance differences exceeding 20\% between the best- and worst-performing subgroups within a single diagnostic category\cite{OakdenRayner2020HiddenStratification}. For pathology, where rare tumor subtypes may constitute only a small fraction of cases but carry substantially different prognoses and treatment implications, this phenomenon poses serious clinical risk. Models deployed without subgroup-specific validation may systematically underserve patients with uncommon presentations, precisely those for whom AI assistance could theoretically provide the greatest benefit\cite{Campanella2025ClinicalBenchmark}.

\hypertarget{supp:B5}{\heading{B.5 Domain Shift and Generalization Failures}}

This section elaborates on the “pervasive domain shift driven by pre-analytic variabilities” highlighted in Main Text Section 3.2 as the most critical aspect of the technical reality gap, providing empirical evidence on its sources, magnitude, and current mitigation strategies.

\hypertarget{supp:B5:DistribShift}{\hheading{Sources of Distribution Shift}}

Digital pathology images exhibit substantial variation across institutions, arising from differences in tissue processing, staining reagents, staining protocols, and whole-slide scanner hardware\cite{otalora2019staining}\cite{Dunn2025stain_variability}. Even identical staining protocols executed at different laboratories produce visibly different images due to variation in reagent batches, incubation times, and environmental conditions. Scanner-induced variation adds another layer: different manufacturers’ devices capture tissue with distinct color responses, compression artifacts, and resolution characteristics.

These variations constitute domain shift: the statistical mismatch between training and deployment data that degrades model performance. As the main text emphasizes, when pretraining datasets lack diversity across these axes, models risk conflating technical batch effects with actual biological signals\cite{schmitt2021hidden}. Studies have shown that stain normalization can improve generalization to external datasets, with reported improvements of 5–11 percentage points in classification accuracy\cite{anghel2019high}. However, stain normalization introduces its own challenges: aggressive normalization can distort diagnostically relevant features, and the choice of reference template influences downstream performance in unpredictable ways\cite{michielli2022stain}.

\hypertarget{supp:B5:MitigationStart}{\hheading{Mitigation Strategies and Their Limitations}}

Multiple approaches have been proposed to address domain shift: stain normalization (template-based color transformation), color augmentation (training with randomized color perturbations), and domain-adversarial training (learning features invariant to domain-specific characteristics)\cite{otalora2019staining}. Tellez, Litjens, and colleagues systematically quantified the effects of these strategies, demonstrating that while stain normalization improves generalization on specific tasks, stain color augmentation during training provides a more robust and computationally efficient alternative that consistently reduces cross-institutional performance degradation\cite{tellez2019quantifying}. Faryna, Litjens, and colleagues further advanced this direction by developing automatic augmentation policies specifically tailored to H\&E-stained histopathology, optimizing augmentation parameters through learned search rather than manual tuning\cite{faryna2021tailoring}. Comparative studies suggest that no single technique dominates across all tasks. Instead, optimal strategies depend on the nature and magnitude of domain shift in specific applications. Foundation models trained on massive multi-institutional datasets\cite{UNI,CHIEF,GigaPath,Virchow} demonstrate improved inherent robustness to domain shift, though residual performance variations across sites persist.

Foundation models pretrained on diverse multi-institutional datasets offer a promising approach to domain generalization by learning representations that abstract away site-specific artifacts. However, even foundation models inherit biases from their pretraining data, and models pretrained predominantly on slides from high-resource academic centers may generalize poorly to resource-constrained settings with different staining practices and scanner equipment. The field lacks systematic characterization of foundation model performance across the full spectrum of clinical deployment environments, a gap that reinforces the need for representations that decouple tissue morphology from the wide spectrum of generation-side variables.

\hypertarget{supp:B6}{\heading{B.6 The Validation Hierarchy}}

\hypertarget{supp:B6:retrospectiveVal}{\hheading{Retrospective Validation: Necessary but Insufficient}}

The overwhelming majority of published computational pathology studies employ retrospective validation designs, evaluating models on held-out portions of curated datasets\cite{ogut2025artificial}. While essential for initial model development and comparison, retrospective validation has well-documented limitations: curated datasets exclude challenging cases that would be encountered in routine practice, labels may reflect the consensus of expert pathologists rather than real-world diagnostic accuracy, and the temporal relationship between model development and evaluation permits subtle data leakage.

Retrospective studies using TCGA or similar research datasets provide evidence of technical feasibility but limited evidence of clinical utility. The gap between retrospective accuracy and prospective clinical performance has been documented across medical imaging domains, with models that appear to match expert performance in controlled evaluation failing to demonstrate benefit in routine clinical use.

\hypertarget{supp:B6:ExternalVal}{\hheading{External Validation: The Geographic and Temporal Imperative}}

Meaningful clinical translation requires validation on data from institutions not involved in model development, namely external validation that tests generalization across the site-specific factors that confound internal evaluation. Stronger evidence comes from geographically diverse validation cohorts: the PANDA challenge for prostate cancer grading, organized by Bulten, Litjens, and colleagues, demonstrated that algorithms achieving 0.86–0.87 agreement with expert pathologists on European and US validation sets represented genuinely cross-continental generalization\cite{PANDA}. As van der Laak, Litjens, and Ciompi noted in their review of clinical implementation barriers, the gap between retrospective validation and prospective clinical utility remains the field’s most persistent translational challenge, requiring careful attention to workflow integration, pathologist acceptance, and continuous monitoring\cite{vanderlaak2021deep}.

Temporal validation which evaluates models on data collected after development completion, provides additional assurance against the phenomenon of concept drift, whereby the relationship between image features and diagnostic labels shifts over time due to evolving clinical practices, new staining reagents, or updated classification systems. Models validated only on contemporaneous data may degrade as clinical practice evolves.

\hypertarget{supp:B6:Prospective}{\hheading{Prospective Validation and Randomized Trials}}

The highest level of clinical evidence requires prospective evaluation of model performance in actual clinical workflows, ideally through randomized controlled trials comparing AI-assisted diagnosis against standard care. Prospective validation assesses not only diagnostic accuracy but integration challenges, workflow disruption, and impact on clinical decision-making which are factors invisible in retrospective analysis. Early prospective studies suggest that AI tools achieving expert-level accuracy in controlled settings may have more modest impact when deployed in routine practice, where time pressures, alert fatigue, and human-AI interaction dynamics attenuate theoretical benefits.

The field faces a validation gap: while numerous AI algorithms have demonstrated expert-comparable accuracy in retrospective evaluation, far fewer have undergone prospective clinical trials, and fewer still have demonstrated improved patient outcomes in randomized studies. This evidence hierarchy (from retrospective to prospective to randomized) maps directly onto the “longitudinal stability, workflow compatibility, and risk management” that the main text identifies as insufficiently captured by current research paradigms\cite{El_Arab2025}\cite{vasey2022reporting}. Regulatory clearance, as discussed in Section B.7, often requires less stringent evidence than would be necessary to demonstrate clinical benefit.

\hypertarget{supp:B7}{\heading{B.7 Regulatory Pathways and Evidence Requirements}}

The regulatory landscape detailed here provides the institutional and legal context for the safety, liability, and risk management challenges discussed in Main Text Section 3.3. The fragmented and evolving nature of these pathways, across the US and EU, illustrates why AI-based pathology systems occupy the “fragmented translational landscape” described in the main text introduction.

\hypertarget{supp:B7:CurrentReg}{\hheading{Current Regulatory Landscape}}

As of early 2026, regulatory pathways for AI-based pathology devices remain nascent relative to radiology, where hundreds of algorithms have received FDA clearance. The FDA approved the first AI-based digital pathology device (PAIGE Prostate) in September 2021 via the De Novo pathway, establishing a new device classification for “software algorithms to provide information to the user about presence, location, and characteristics of areas of the image with clinical implications”\cite{petrick2023regulatory}. Subsequent clearances have followed, including the recent 510(k) clearance of Ibex Prostate Detect for detecting small and rare prostate cancers, which demonstrated 99.6\% positive predictive value for heatmap accuracy and identified 13\% of cancer cases missed by pathologists on initial reads\cite{pantanowitz2020ai_prostate}\cite{santarosario2024ibex}.

The European regulatory landscape differs in both structure and evidence requirements. CE marking under the In Vitro Diagnostic Regulation (IVDR) mandates conformity assessment by notified bodies, with requirements varying by device risk class. Notably, only approximately half of digital pathology AI products with published evidence had their publications appear before regulatory approval, suggesting that market access often precedes peer-reviewed performance validation\cite{matthews2024public_evidence}.

\hypertarget{supp:B7:EvidenceGap}{\hheading{Evidence Gaps and Emerging Standards}}

Analysis of FDA-cleared AI medical devices reveals that only 28\% reported multi-site evaluation, and only 3\% used prospective data\cite{matthews2024public_evidence}. This evidence gap creates a paradox: regulatory clearance enables market access but does not guarantee clinical adoption, reimbursement, or demonstrable patient benefit, which directly reinforcing the “lack of sustainable reimbursement pathways” discussed in Main Text Section 3.1. The FDA’s 510(k) pathway, which permits clearance based on substantial equivalence to predicate devices, has facilitated market entry for many AI technologies but sets a lower evidentiary bar than the more rigorous Premarket Approval (PMA) process reserved for highest-risk devices.

Emerging regulatory innovations may help bridge this gap. The FDA’s Biomarker Qualification Program recently qualified PathAI’s AIM-MASH AI Assist as the first AI-powered pathology tool to receive FDA qualification, for use in clinical trials assessing metabolic dysfunction-associated steatohepatitis\cite{pulaski2025mash_validation}\cite{Iyer2024-kg}. This qualification, following similar EMA qualification earlier in 2025, demonstrates that AI-assisted histology can meet regulatory standards for clinical trial endpoints when supported by extensive analytical and clinical validation. The pathway from drug development tool to routine diagnostic use remains to be established\cite{ema2025mash_qualification}.

\endgroup

\setcounter{figure}{0}
\renewcommand{\thefigure}{S\arabic{figure}}
\begin{figure*}
\setstretch{0.9}
\centering
\includegraphics[width=0.9\textwidth]{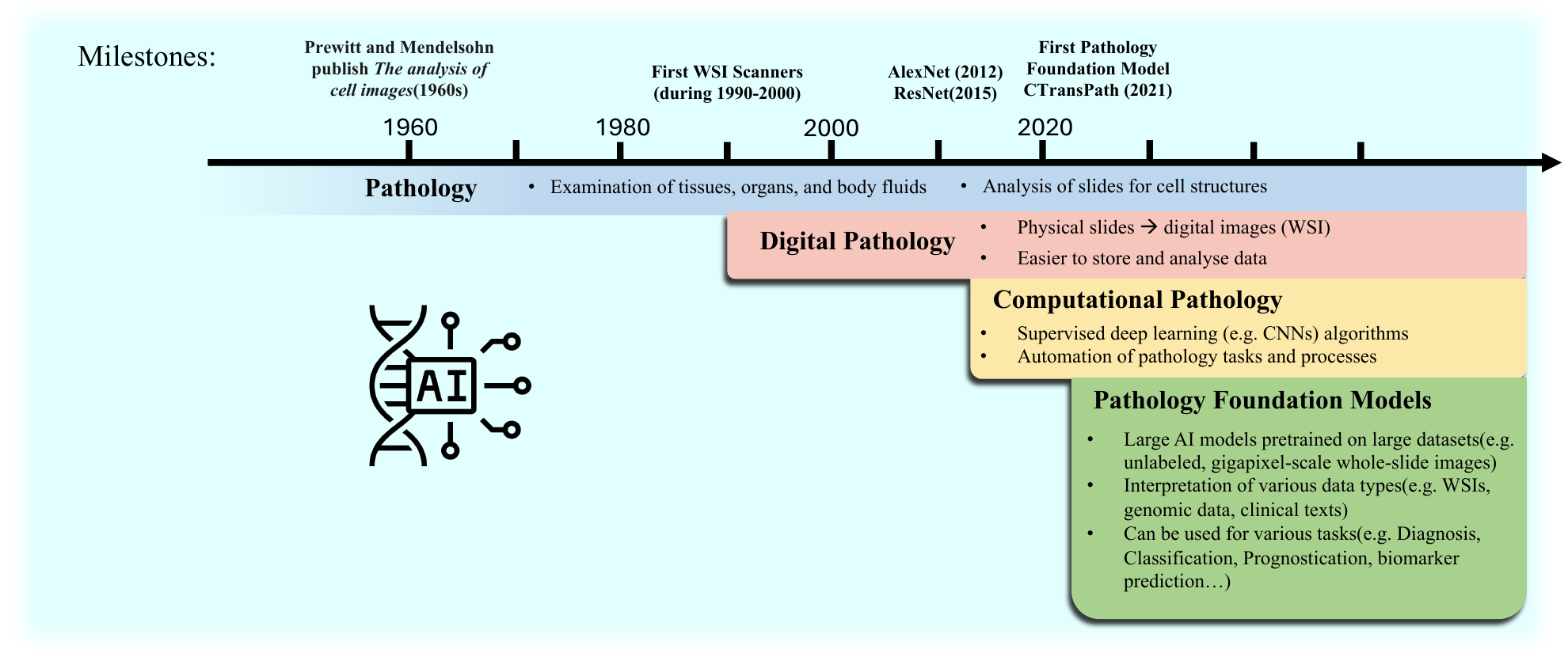}
\caption{Timeline of digital pathology milestones \PC{illustrating key developments from traditional microscopic analysis to pathology foundation models}. The 1960's marked the beginning of computerized microscopy image analysis for disease research \PC{through automated} examinination of cellular structures. The 1980s saw the introduction of whole slide imaging (WSI)\cite{STAMP} scanners, \PC{enabling the conversion of physical slides into digital images and facilitating} data storage and analysis. In the 2000s, commercial WSI scanners became widely available, \PC{with AI algorithms beginning to automate pathology tasks through digital image analysis}. By the 2020s, AI-based medical devices leveraging CNNs \cite{2019Wang_S_TheAmericanJournal_CNNSeg} were developed, \PC{enabling diverse data type interpretation and multi-task support}, further advancing the automation of disease diagnosis and clinical decision-making. The \PC{emergence of pathology foundation models represents the current frontier}, moving the field toward comprehensive automation and intelligent tissue analysis.}
\label{fig:pathologytime}
\end{figure*}

\begin{table*}
\footnotesize
\centering
  \caption{\setstretch{0.9} \PC{Pathology foundation models with their key characteristics and specifications. ViT stands for Vision Transformer, with model sizes including ViT-Base (ViT-B), ViT-Large (ViT-L), ViT-Huge (ViT-H), and ViT-Giant (ViT-G), which refer to increasing parameter counts and model complexity. The table summarizes publication dates}, dataset scales, magnification levels, research sources, model architectures, training strategies, and computational \PC{requirements}. These models leverage advanced machine learning techniques, including self-supervised methods like DINOv2 and iBOT, as well as multimodal approaches integrating both image and text data. Dataset scales range from thousands to millions of images, with magnifications typically between 10× and 40×. \PC{Training utilizes large-scale hardware configurations including A100 and V100 GPUs to support the computational demands of these architectures}.} \label{table:list of model}
    \resizebox{\textwidth}{!}{
    \begin{tabular}{
    >{\columncolor{headerblue}\centering\arraybackslash}p{2cm}
    >{\columncolor{headerblue}\centering\arraybackslash}p{1.5cm}
    >{\columncolor{headerblue}\centering\arraybackslash}p{4.5cm}
    >{\columncolor{headerblue}\centering\arraybackslash}p{2cm}
    >{\columncolor{headerblue}\centering\arraybackslash}p{3cm}
    >{\columncolor{headerblue}\centering\arraybackslash}p{4cm}
    >{\columncolor{headerblue}\centering\arraybackslash}p{1.5cm}
    >{\columncolor{headerblue}\centering\arraybackslash}p{3cm}
    >{\columncolor{headerblue}\centering\arraybackslash}p{3cm}
    >{\columncolor{headerblue}\centering\arraybackslash}p{1cm}
    }
    \toprule
    \textbf{Model} & \textbf{Released} & \textbf{Dataset Scale} & \textbf{Magnification} & \textbf{R\&D Source} & \textbf{Architecture} & \textbf{Training} & \textbf{Hardware} & \textbf{Modality} & \textbf{Slide Enco} \\
    \midrule
    \rowcolor{rowgray}
        RetCCL & 2021.Dec & 32K WSIs, 15M Patches & 5×,20× & TCGA,PAIP & ResNet50 & MoCov2 & 32 V100 32G GPUs & Single Modality & no \\
        CTransPath & 2022.May & 32.2k WSIs, 15.6M Tiles & 20×, 40× & TCGA,PAIP & Swin Transformer & MoCoV3 & 48 V100 GPUs & Single Modality & no \\
    \rowcolor{rowgray}
        HIPT & 2022.Jun & 10K WSIs, 104M Patches & 20× & TCGA & ViT-S/16 & DINO & -- & Single Modality & no \\
        REMEDIS & 2022.Jul &  29K WSIs, 50M Patches & 20× & TCGA & ResNet-152(2×) & SimCLR & -- & Single Modality & no \\
    \rowcolor{rowgray}
        LunitDINO & 2023.Apr & 36K WSIs, 32M Patches & 20×, 40× & TCGA, TULIP &  ViT-S/(8,16) & DINO & 64 V100 GPUs & Single Modality & no \\
        Phikon & 2023.Jul & 6K WSIs, 43M Patches & 20× & TCGA & ViT-B/16 & iBOT & 64 V100 GPUs & Single Modality & no \\
    \rowcolor{rowgray}
        PLIP & 2023.Aug & 208K Image-Caption Pairs & -- & OpenPath & ViT-B/32(image), Masked Transformer(text) & CLIP & -- & Multimodal & no \\
        Kaiko & 2024.Mar & 29K WSIs, 256M Patches & 5×, 10×, 20×, 40× & TCGA & ViT-S, ViT-B, Vit-L & DINOv2 &  16 H100 GPUs & Single Modality & no \\
    \rowcolor{rowgray}
        PathAsst & 2024.Mar & 207K Tile-Caption Pairs & -- & PathCap, PathInstruct & PathCLIP, Vicuna-13B & CLIP & -- & Multimodal & no \\
        CONCH & 2024.Mar & 1170K Images-Descriptions & 10×, 20× & PubMed Central-OA & ViT-B/16(visual), L12-E768-H12(text) & CoCa & 8  A100-80G GPUs & Multimodal & no \\
    \rowcolor{rowgray}
        UNI & 2024.Mar & 100K WSIs & 20×, 40× & Mass-100K(MGH, Brigham, BWH) & ViT-L/16 & DINOv2 & 32 A100-80G GPUs & Single Modality & no \\
        PLUTO & 2024.May & 158K WSIs, 195M Images & 20×, 40× & TCGA, Private dataset, and more & FlexiViT-S & DINOv2 (iBOT) & 64 A40 GPUs & Single Modality & no \\
    \rowcolor{rowgray}
        Prov-GigaPath & 2024.May & 171K WSIs, 1.3B Images & 20× & Providence & ViT-H/14, LongNet & modified DINOv2 & 64 A100-80G GPUs & Multimodal & yes \\
        PRISM & 2024.May & 587K WSIs, 197K relevant reports & 20× & Private dataset & PerceiverNet \& Virchow(image), BioGPT(text) & CoCa & 16 V100-32G GPU & Multimodal & yes \\
    \rowcolor{rowgray}
        PathChat & 2024.May & 1180K Image-Caption Pairs & -- & PubMed Central-OA & UNI(image), Llama2(text) & LLaVA & 8 80GB GPUs & Multimodal & no \\
        TANGLE & 2024.May & 47K WSIs, 6.5K Image-Gene Pairs & 20× & TG-GATEs & ViT-B/16, ABMIL & iBOT & -- & Single Modality & yes \\
    \rowcolor{rowgray}
        RudolfV & 2024.Jun & 133K WSIs, 1250M Patches & -- & TCGA, In-house & ViT-L/14 & DINOv2 & 16 A100-40GB GPUs & Single Modality & no \\
        HIBOU-B & 2024.Jun & 1.1M WSIs, 512M Patches & 20× & Private dataset & ViT-B/14 & DINOv2 & 8 A100-80G GPUs & Single Modality & no \\
    \rowcolor{rowgray}
        HIBOU-L & 2024.Jun & 1.1M WSIs, 1.2B Patches & 20× & Private dataset & ViT-L/14 & DINOv2 & 32 A100-40G GPUs & Single Modality & no \\
        BEPH & 2024.Jun & 11K WSIs, 11M Patches & 20× & TCGA & Vit-B/16(VQ-KD) & BEiTv2 & 8 A100 GPUs & Single Modality & no \\
    \rowcolor{rowgray}
        GPFM & 2024.Jul & 72K WSIs, 190M Patches & -- & TCGA, GTEx, and more & Vit-L & modified DINOv2 & 2×8 80GB H800 GPUs & Single Modality & no \\
        mSTAR+ & 2024.Jul & 11K WSIs, 22K Modality Pairs & 20× & TCGA & ViT-L & CLIP & 4 × 80 GB H800 GPUs & Multimodal & no \\
    \rowcolor{rowgray}
        H-optimus-0 & 2024.Jul & 500K WSIs & 20× & Private dataset & ViT-G & DINOv2 / iBOT & 8 × A100-80G GPUs & Single Modality & yes \\
        Virchow & 2024.Jul & 1,500K WSIs & 20× & MSKCC & ViT-H/14 & DINOv2 & 16 V100 GPUs & Single Modality & no \\
    \rowcolor{rowgray}
        Pathoduet & 2024.Jul & 14K WSIs & -- & TCGA, HyReCo, BCI & ViT-B & MoCoV3 extension & 8 A100 GPUs & Single Modality & yes \\
        CHIEF & 2024.Aug & 60K WSIs (semi-supervised), 15M WSIs (self-supervised) & 10× & Proprietary dataset(DFCI, BWH, YH, SMCH, CUCH) & CTransPath(image), Transformer Layer(text) & CLIP & 8 V100-32G GPUs & Multimodal & yes \\
    \rowcolor{rowgray}
        Phikon-v2 & 2024.Sep & 58K WSIs & 5×, 20× & PANCAN-XL & ViT-L & DINOv2 & 4 32GB V100 GPUs & Single Modality & no \\
        MADELEINE & 2024.Oct & 16K WSIs & 10×, 20×, 40× & Acrobat, BWH & CONCH, MH-ABMIL & CLIP, GOT & 3 × 24GB 3090Ti GPUs & Single Modality & yes \\
    \rowcolor{rowgray}
        Virchow2 & 2024.Nov & 
        3.1M WSIs & 5×, 10×, 20×, 40× & MSKCC, Private dataset & ViT-H/14 & DINOv2 & 512 Nvidia V100 GPUs & Single Modality  & yes \\
        TITAN & 2024.Nov & 335K WSIs & 20× & Mass-340K & iBOT-ViT(image), CoCa-LLM(text) & CoCa & 8 A100-80G GPUs & Multimodal & no \\
    \rowcolor{rowgray}
        MINIM & 2024.Dec & 59K+74K Image-Text Pairs; 30K+79K Tumor-Annotated Images & -- & Private dataset & Diffusion model, BERT tokenizer & U-Net, CLIP & 8 A100 GPUs & Multimodal & no \\
        KEEP & 2024.Dec & 143K Image-Text Pairs & -- & Quilt-1M, OpenPath & UNI & CLIP & 4 A100 GPUs & Multimodal & no \\
    \rowcolor{rowgray}
        MUSK & 2025.Jan & 32K WSIs, 1B Text Tokens, 1M Image–Text Pairs & 10×, 20×, 40× & PMC OA, TCGA, QUILT-1M & MoE in LLM(V-FFN, L-FFN) & modified CoCa, BEiT-3 & - & Multimodal & no \\
        THREADS & 2025.Jan & 47K WSIs, 125M Patches, 26K Bulk RNA, 20K DNA Variants & 20×, 40× & MGH, TCGA, and more & CONCHv1.5 & CLIP & 4 × 80GB A100 GPUs & Multimodal & yes \\
    \rowcolor{rowgray}
        Atlas & 2025.Jan & 1.2M WSIs, 3.4B Patches & 5×, 10×, 20×, 40× & Proprietary dataset(Mayo Clinic, Charité - Universtätsmedizin Berlin) & ViT-H/14 & DINOv2 & H100 GPU & Single Modality & yes \\
        UNI2-h & 2025.Jan & 350K WSIs, 200M Images & --- & Mass General Brigham & ViT-h/14-reg8 & DINOv2 (iBOT) & lots of A100 80GB GPUs & Single Modality & - \\
    \rowcolor{rowgray}
        OmiCLIP & 2025.May & 2.2M Patches-Transcriptomic Data Pairs & -- & ST-bank & ViT, Causal Masking Transformer & CoCa & 1 A100 80-GB GPU & Multimodal & no \\
      \bottomrule
    \end{tabular}%
}
\end{table*}